\documentclass[10pt]{article}

\usepackage{amsmath}
\usepackage{amssymb}

\usepackage{breqn}

\usepackage{graphicx}

\usepackage{cite}

\usepackage{color} 
\usepackage{soul}

\usepackage{textcomp}

\usepackage[labelfont=bf,labelsep=period,justification=raggedright]{caption}

\makeatletter
\renewcommand{\@biblabel}[1]{\quad#1.}
\makeatother

\date{}

\begin{document}

\begin{flushleft}
{\Large
\textbf{Bayesian Decision Tree for the Classification of the Mode of Motion in Single-Molecule Trajectories}
}
\\
Silvan T\"urkcan$^{1,2,3\ast }$, 
Jean-Baptiste Masson$^{1,2}$,
\\
\bf{1}  Physics of Biological Systems, Institut Pasteur, Paris, France
\\
\bf{2}  Centre National de la Recherche Scientifique (CNRS), UMR 3525, Paris, France
\\
\bf{3}  Laboratoire d'Optique et Biosciences, Ecole Polytechnique, Centre National de la Recherche Scientifique, Institut National de la Sant\'{e} et de la Recherche M\'{e}dicale U696, Palaiseau, France
\\
$\ast$ E-mail: silvan@stanford.edu
\end{flushleft}

\section*{Abstract}

Membrane proteins move in heterogeneous environments with spatially (sometimes temporally) varying friction and with biochemical interactions with various partners. It is important to reliably distinguish different modes of motion to improve our knowledge of the membrane architecture and to understand the nature of interactions between membrane proteins and their environments. Here, we present an analysis technique for single molecule tracking (SMT) trajectories that can determine the preferred model of motion that best matches observed trajectories. The method is based on Bayesian inference to calculate the posteriori probability of an observed trajectory according to a certain model. Information theory criteria, such as the Bayesian information criterion (BIC), the Akaike information criterion (AIC), and modified AIC (AICc), are used to select the preferred model. The considered group of models includes free Brownian motion, and confined motion in $2$nd or $4$th order potentials. We determine the best information criteria for classifying trajectories. We tested its limits through simulations matching large sets of experimental conditions and we built a decision tree. This decision tree first uses the BIC to distinguish between free Brownian motion and confined motion. In a second step, it classifies the confining potential further using the AIC. We apply the method to experimental Clostridium Perfingens $\epsilon$-toxin (CP$\epsilon$T) receptor trajectories to show that these receptors are confined by a spring-like potential. An adaptation of this technique was applied on a sliding window in the temporal dimension along the trajectory. We applied this adaptation to experimental CP$\epsilon$T trajectories that lose confinement due to disaggregation of confining domains. This new technique adds another dimension to the discussion of SMT data. The mode of motion of a receptor might hold more biologically relevant information than the diffusion coefficient or domain size and may be a better tool to classify and compare different SMT experiments.\\

%

\section*{Introduction}

\begin{figure}[!ht]
\begin{center}
\includegraphics[width=3.27in]{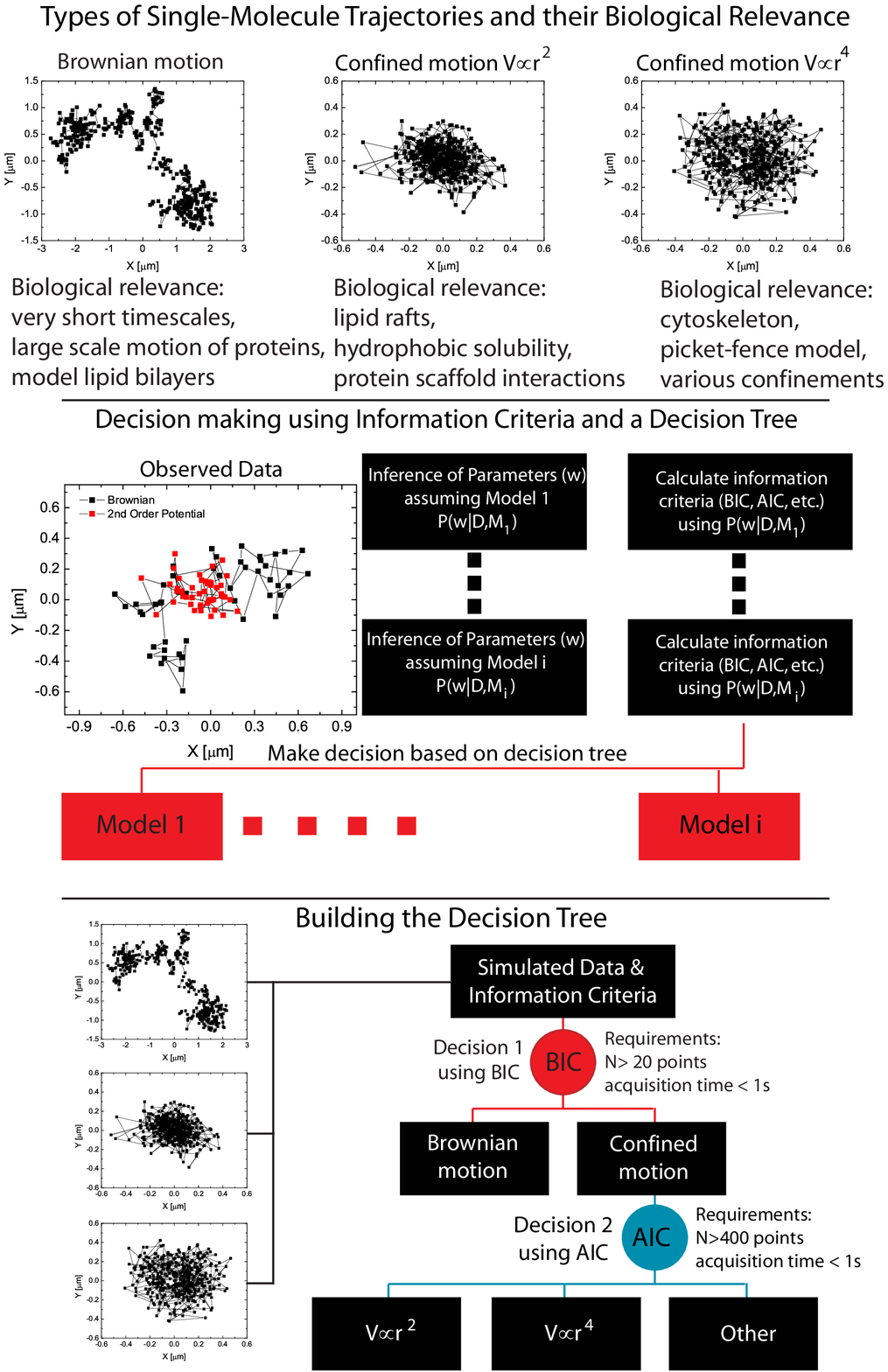}
\end{center}
\caption {
{\bf Bayesian Decision Tree for the Classification of Single-Molecule Trajectories.} Biomolecules undergo a variety of different modes of motion in the cell membrane, which are often difficult to distinguish. We show Brownian motion, and confined motion in a 2nd and 4th order potential as examples for receptors that might reside in lipid rafts or move according to the picket-fence model. Using a Bayesian inference and a decision tree, which can be developed through simulations with known modes of motion, it is possible to easily classify modes of motion of molecules in the cell membrane. Adapted decision criteria, such as the Bayesian information criterion (BIC) or the Akaike information criterion (AIK) can be computed from the maximum a posteriori distribution (MAP) and used to make decisions on the single-trajectory level. The decision tree that was derived for this work is shown in the bottom of the figure. We first use the BIC (red) to determine if a potential confines the biomolecule and then classify the type of potential using the AIC (blue).
}
\label{Fig1}
\end{figure}

Advances in single molecule tracking (SMT) techniques, have made it possible to record trajectories of individual biomolecules in a large variety of biological systems \cite{Saxton1997,lord2010single}. This allows for new insight into the dynamics of membrane proteins and into the structural organization of the membrane. Labeled membrane biomolecules can undergo free Brownian diffusion, confined motion, hopping, stabilization by scaffolding proteins, anomalous diffusion etc. The complex motion of membrane proteins has been attributed to molecular crowding effects \cite{ryan1988molecular,Dix2008}, intermolecular interactions \cite{Sieber2006,douglass2005single}, differences in lipid solubility \cite{turkcan2013probing}, cytoskeleton barriers \cite{sheetz1993glycoprotein,Kusumi2005}, non-local potential fields induced by the environment \cite{masson2009inferring, turkcan2012exp,turkcan2012bayesian}, tethering to the cytoskeleton \cite{Peters1999,Jin2007}, lipid rafts or domains \cite{Varma1998,lingwood2010lipid} and hopping between confinement areas\cite{turkcan2013probing}. Finally, proteins often exhibit a mix between these behaviors that lead to different modes of motion (Fig. \ref{Fig1} top).\\

It is important to reliably distinguish between different modes of motion of molecules and to quantify their characteristics. This allows to gain deeper insights into the structure of the membrane and to better understand the nature of the interactions between proteins and their environments. The most widely spread approach to classify the mode of motion is based on the analysis of the mean-square displacement (MSD) of the tracked molecule \cite{Barak1982,Sheetz1989}. The MSD is usually plotted against the time lag $\tau$. In the case of Brownian motion, the resulting points should lie on a line, whose slope is proportional to the diffusion coefficient $D$ (for $2$ dimensional Brownian diffusion $MSD(\tau)=4D\tau^{\alpha}, \alpha = 1$). If the relationship is not linear, the motion of the molecule is classified as subdiffusive ($\alpha < 1$) or superdiffusive ($\alpha > 1$). In the case of confined motion, the particle does not escape from a corral of a certain size during the observed time, which will manifest itself in the MSD versus time lag plot through a plateau. Yet, this method is known to fail to take into account diffusion heterogeneities and transient confinement may be misinterpreted as anomalous diffusion. Hence, it tends to often identify biomolecule motion as subdiffusive and leads to extremely wide distributions of diffusion coefficients that are difficult to associate to physical characteristics of a medium. A different approach is to analyze the cumulative distribution of square displacements for a fixed lag time. Here, the cumulative distribution is analyzed for individual trajectories \cite{Pinaud2009} or multiple trajectories \cite{schutz1997,Deverall2005}. Similarly, this method is sensitive to heterogeneities and does not include interactions with the local environment. Image correlation techniques are also used \cite{hebert2005spatiotemporal}, as well as a technique based in Bayesian inference \cite{monnier2012bayesian}. Other methods exploit higher-order moments of the biomolecule displacement \cite{coscoy2007statistical}, first passage times \cite{condamin2008probing} and the analysis of the radial density distribution \cite{Jin2007}. These estimators also tend to introduce the heterogeneities into the quantity describing the motion or are sensitive to local geometrical effects. Finally, all of these methods exploit only a subset of the available information because they either discard part of the full information or loose information through averaging. \\

A related challenge is to correctly subdivide trajectories that undergo different modes of motion and to clearly determine when the mode of motion changed \cite{Kusumi2005,burckhardt2009virus}. Receptors have been observed to be transiently confined in small corrals and to then continue with free Brownian diffusion \cite{Pinaud2009}. One method to detect transient confinement is to evaluate the probability that the tracked molecule will stay within a certain region during a short window that is moving along the entire trajectory \cite{saxton1993lateral,simson1995detection}. Yet, this method fails when the environment is highly heterogeneous or when local interactions are sufficiently strong to significantly deviate the trajectories from pure Brownian motion. A more versatile method evaluates the diffusion coefficient, the MSD curvature, and trajectory asymmetry with a threshold for a variable window. It has been shown to segment trajectories into portions exhibiting stalled, constrained, directed or Brownian motion \cite{huet2006analysis}. Different types of motion can also be identified by a characteristic feature vector \cite{helmuth2007novel}. Features that can distinguish between directed motion, fast and slow drifting motion, and confined motion are: net displacement, straightness, bending, efficiency, asymmetry and skewness. Another wavelet transform based method can reliably detect dynamic heterogeneities in time series data without making prior assumptions about a model, but it does not give information on what model best describes the data \cite{chen2013diagnosing}.\\ 

In an attempt to address the high heterogeneity of the membrane, to take the local and non-local interactions into account and to exploit as much information as possible from SMTs, we previously developed a Bayesian inference scheme allowing the spatial mapping of both the diffusivity and the potential field \cite{turkcan2012bayesian,masson2009inferring,turkcan2013probing}. Here, we extend the Bayesian scheme to classify the nature of protein motion in the membrane by building a solid tool for discriminating between modes of motion. The studied cases are free Brownian motion, the harmonic confined trajectories $V\propto r^{2}$ and the anharmonic confined trajectories $V\propto r^{4}$. These 3 simple models cover a large set of possible membrane molecule behaviors. Confinement can stem from a large variety of interactions, ranging from purely local interactions, for example with the cytoskeleton, to highly non local interactions, for example complex organization of different lipids. The higher the anharmonicity of the potential, the more it can model localized interactions. The here presented classification technique is an extension of the inference technique that can provide a comprehensive measure of the fit of a used model. The previously introduced inference technique only returns the most likely values for the model parameters, but it does not provide a relative value for the goodness of fit of the chosen model over alternative models.  This quantitative model comparison is an additional layer of analysis that is added after the previously introduced inference method is used for each model separately.\\

To classify the motion, we infer the parameters that best fit an observed trajectory, assuming a model for the motion and obtain the maximum a posteriori (MAP). Using the MAP, we can calculate various information based criteria, such as the Bayesian information criterion (BIC) and the Akaike information criterion (AIC) for each tested model and use them to determine the preferred model (Figure \ref{Fig1} middle). These criteria are often better suited for model selection than the direct likelihood-ratio test, because they take the degrees of freedom of the model into account. Moreover, to test the accuracy and to further enhance the preferred model selection, we built a decision tree using simulated data for all models (Figure \ref{Fig1} bottom). We then developed a variant of the technique that performs the model selection with a temporal resolution. This provides a tool that can detect changes in the mode of motion of a biomolecule.\\

Extensive simulations with numerical trajectories undergoing Brownian motion in confining potentials are used to build a decision tree. The decision tree first determines if a trajectory is undergoing free Brownian motion or if it is confined by a potential through the Bayesian information criterion. If the trajectory is classified as being confined by a potential, the decision tree can further distinguish between potentials of the type $V\propto r^{2}$ and $V\propto r^{4}$. Note that there are no theoretical limitations imposing the analysis to specifically stop at order $4$ for the potential. Yet, often experimental recordings are limited to few hundred points, hence the inference for higher order potentials would not gather significantly more information. Then, we apply this technique to experimental trajectories of the \textit{Clostridium Perfingens} $\epsilon$-toxin (CP$\epsilon$T) receptor on live Madin-Darby Canine Kidney (MDCK) cell membranes and confirm that these receptors are confined in a harmonic-like potential, as previously shown \cite{turkcan2012exp}. Furthermore, we apply the temporal version of the model selection to experimental CP$\epsilon$T receptor trajectories that lose confinement due to disaggregation of confining domains. The CP$\epsilon$T is a member of the pore-forming toxin family. It is secreted by the bacterium as prototoxin monomers and activated by an enzymatic reaction. An individual toxin binds to a currently unidentified 37-kDa membrane receptor on the cell membrane, oligomerizes with other bound toxins, and forms pores that pierce the cell membrane with a $\beta$-barrel and cause the death of the cell by uncontrolled ion exchange \cite{petit1997clostridium,Tilley2006}.\\

\section*{Results}

\subsection*{Information theory criteria}

To determine the preferred model corresponding to a biomolecule trajectory, we calculate the Bayesian information criterion (BIC), the Akaike information criterion (AIC), and a modified version of the AIC whose performance improves for short trajectories, the corrected AIC (AICc). These criteria were developed to determine the preferred model, or the model that best describes a given data set. Note that to build the decision tree, trajectories were generated from the $3$ predefined models. Experimentally, trajectories will be associated to one of the $3$ models which describe a large set of experimental situations, yet it is obviously possible to extend the decision tree by adding other types of motion.\\

The AIC is based on Kullback-Leibler information loss \cite{burnham2004} and can be understood as information entropy. The AIC is given by

\begin{eqnarray}
AIC=-2\cdot\ln(L)+2k \label{AIC}
\end{eqnarray}
\\

where k is the number of free parameters in the model and L is the MAP of the likelihood function assuming a certain model. Note that in our case the number of parameters greatly varies between models with $1$ variable for the Brownian motion, $6$ variables for harmonic confinement and $16$ variables for anharmonic $4$th order confinement. The preferred model is the one that yields the lowest AIC value, which is a measure of lost information. The AICc is a variant of the AIC with a correction for finite sample sizes $N$ and is given by 

\begin{eqnarray}
AICc=AIC+\frac{2k(k+1)}{N-k-1} \label{AICc}
\end{eqnarray}
\\

The BIC, also known as the Schwarz criterion, is based on Bayes factors and is derived from the asymptotic behavior of Bayes estimators under a special class of priors \cite{Schwarz}. The preferred model is the model that yields the minimal value for:

\begin{eqnarray}
BIC=-2\cdot\ln(L)+k\cdot \ln(N) \label{BIC}
\end{eqnarray}
\\

These three information criteria are calculated for each individual trajectory and then a decision about the preferred model is made according to each criterion. Such an approach is more adapted to single-molecule data. Averaging values of the information criteria over many trajectories will lead to overlapping distributions and to the inability to decide between different models, as shown in figure S1. Furthermore, the high heterogeneity of biological media prevent the use of averaging between different spatial parts of the cell.\\

We will next evaluate how many numerical trajectories are correctly attributed to their respective mode of motion under various conditions matching the ones met experimentally. This will lead to a multidimensional map that will give us the experimental parameter range for which this technique is valid. Furthermore, we will use these results to build the decision tree. Here, the studied cases are free Brownian motion, the harmonic confined trajectories $V\propto r^{2}$ and the anharmonic confined trajectories $V\propto r^{4}$. Specifically, we use the 2nd order potential $V(x,y)=1/2k_{X}x^2+1/2k_{Y}y^2$ with $k=\sqrt{k^{2}_{X}+k^{2}_{Y}}$ and the 4th order potential $V(x,y)=\alpha_{X}x^4+\alpha_{Y}y^4$ with $\alpha=\sqrt{\alpha^{2}_{X}+\alpha^{2}_{Y}}$. These 3 simple models cover a large set of possible membrane molecule behaviors. We will not discuss diffusion heterogeneities, since they were treated in previous work \cite{turkcan2012bayesian,turkcan2012exp,turkcan2013probing}. Indeed, free Brownian motion in heterogeneous diffusive environments can be modeled and inferred as free Brownian motion with a locally varying diffusion coefficient $D(r)$.\\

\subsection*{Free Brownian motion}

\begin{figure}[!ht]
\begin{center}
\includegraphics[width=3.27in]{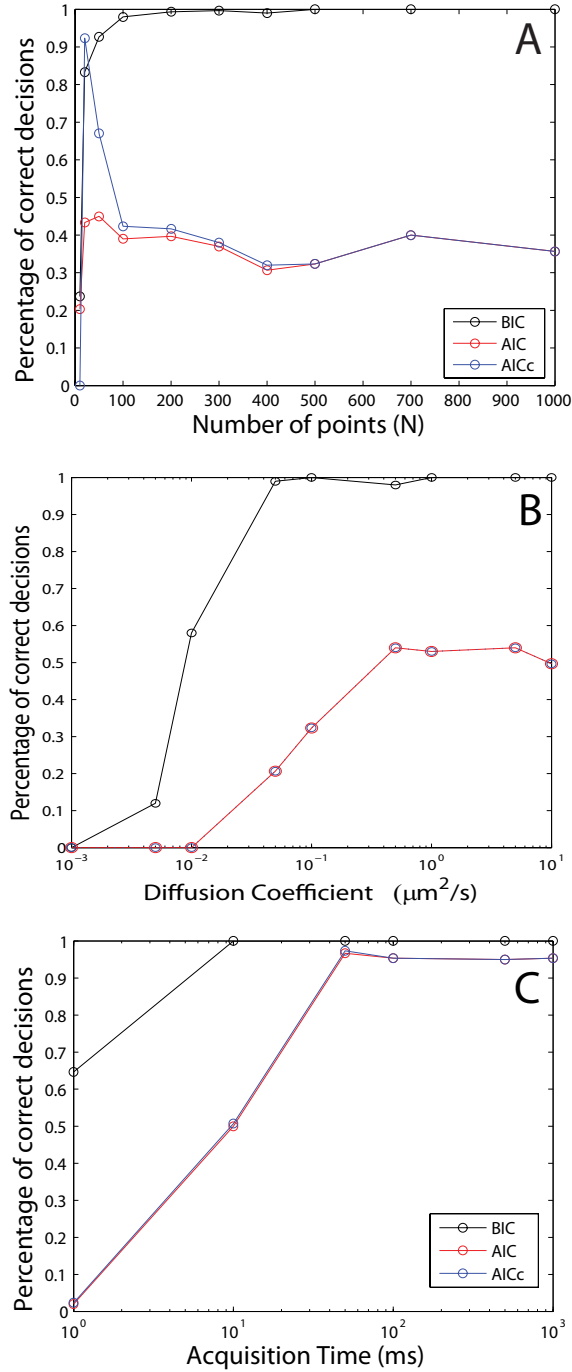}
\end{center}
\caption{
{\bf Information criteria for simulated free Brownian trajectories.} To determine the performance of the decision criteria, we calculated the BIC (black), AIC (blue) and AICc (red) for simulated Brownian trajectories under various conditions. (A) Percentage of correct decisions ($300\,$trajectories per point) versus the length of the trajectory (Parameters: $D_{input}=0.1\,\mu m^{2}/s$, $t_{acq}=50\,ms$, $B_{r}=30\,nm$). The BIC outperforms the AIC and AICc. (B) Percentage of correct decisions versus the input diffusion coefficient (Parameters: N$=500\,$points, $t_{acq}=50\,ms$, $B_{r}=30\,nm$). The BIC outperforms the AIC and AICc and works down to a diffusion coefficient of $0.01\,\mu m^{2}/s$. (C) Percentage of correct decisions versus acquisition time (Parameters: $D_{input}=0.1\,\mu m^{2}/s$, $N=500\,$points, $B_{r}=30\,nm$). The BIC outperforms the AIC and AICc and works for acquisition times between $10\,ms$ and $1000\,ms$.
}
\label{Fig2}
\end{figure}

Here, we determine how well the three information theory criteria attribute the correct model to numerical free Brownian trajectories. We studied the performance over a range of trajectory lengths (\textit{N}), diffusion coefficients (\textit{D}), and acquisition times ($t_{acq}$) matching most of the experimental conditions and most of biological media properties. Experimentally, the $t_{acq}$ is the time over which the camera integrates the arriving photons plus the readout time and images are acquired back to back. In simulations $t_{acq}$ is the time between two simulated points. Under each condition, which is specified in the figure caption, we simulated $300$ trajectories and recorded the normalized histogram of decisions, which are made after each trajectory by determining which model yielded the minimal value for a criterion. The performance of the BIC, AIC and AICc with respect to the trajectory length is shown in figure \ref{Fig2} (A). The BIC clearly outperforms the AIC and the AICc, even for very short trajectories. Only $20\,$ points are sufficient for the BIC to correctly find the model of motion, while a length of $10$ points is too short. Fig. \ref{Fig2} (B), shows the range of diffusion coefficients, for which the criteria perform well. Again, the BIC outperforms the AIC and AICc in accuracy, as well as range. Under the specified conditions, the BIC reliably attributed the correct model to the simulated trajectories for a \textit{D} between $0.01\,\mu m^{2}/s$ and $5\,\mu m^{2}/s$. This is the range of diffusion coefficients that is typically observed for membrane molecules. The BIC also outperforms the AIC and AICc with respect to the possible acquisition times. When the input \textit{D} is $0.1\,\mu m^{2}/s$, the BIC is reliable for acquisition times between $10$ and $200\,ms$. These data indicate that the BIC is the better indicator for free Brownian motion, when compared to the AIC and AICc. It should be noted that this evaluation is solely for the model of the trajectory and not the inferred parameters, which might be subject to a bias, as discussed in reference \cite{turkcan2012bayesian}.\\

\subsection*{Confined motion in a harmonic potential ($V\propto r^{2}$)}

\begin{figure}[!ht]
\begin{center}
\includegraphics[width=4.86in]{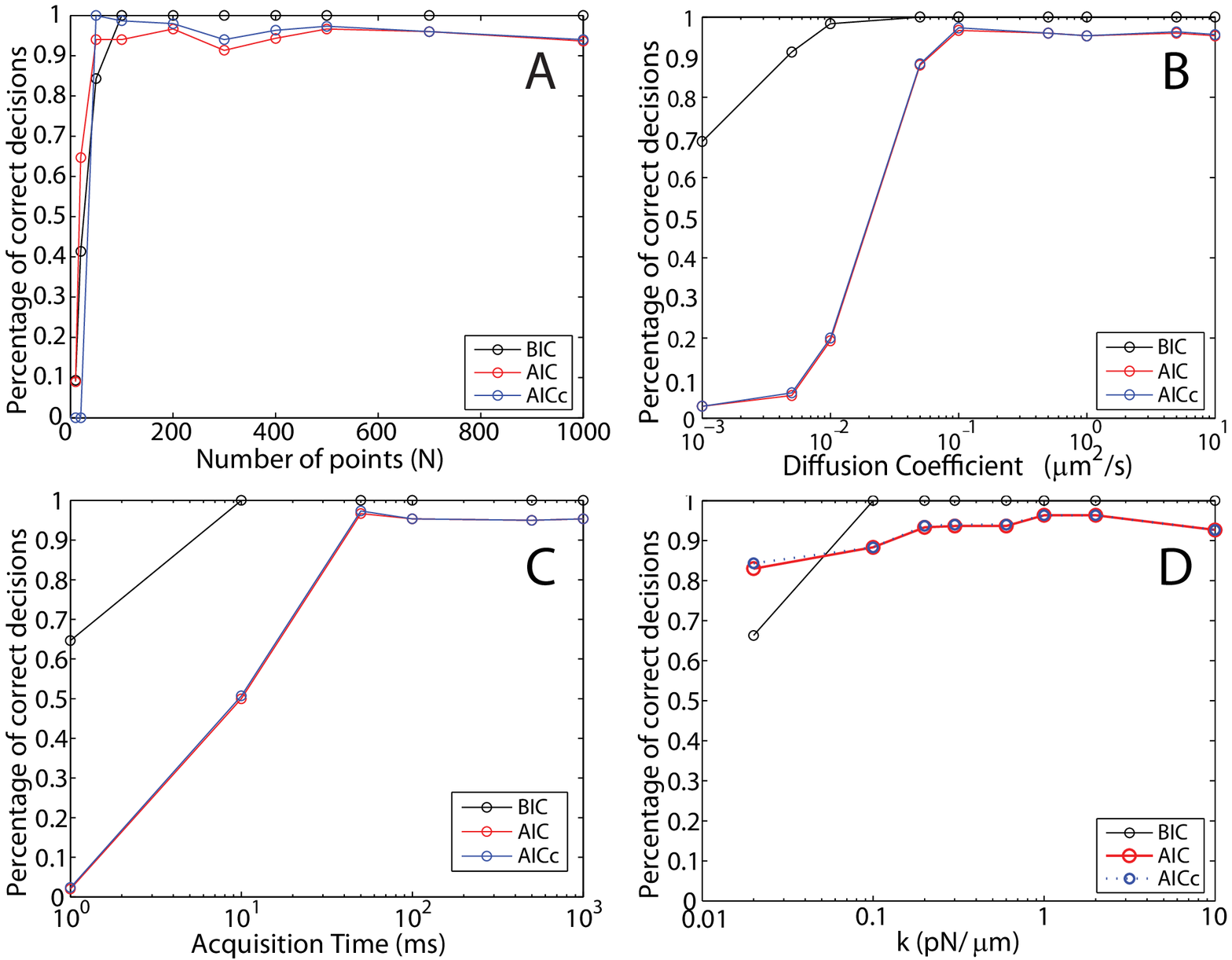}
\end{center}
\caption{
{\bf Information criteria for simulated Brownian trajectories confined in a spring-like potential ($V=1/2kr^{2}$).} To determine the performance of the decision criteria, we calculated the BIC (black), AIC (blue) and AICc (red) for trajectories under various conditions. (A) Percentage of correct decisions ($300\,$trajectories per point) versus the length of the trajectory (Parameters: $D_{input}=0.1\,\mu m^{2}/s$, $t_{acq}=50\,ms$, $B_{r}=30\,nm$, $k=0.3\,pN/\mu m$). The BIC outperforms the AIC and AICc. (B) Percentage of correct decisions versus the input diffusion coefficient (Parameters: N$=500\,$points, $t_{acq}=50\,ms$, $B_{r}=30\,nm$, $k=0.3\,pN/\mu m$). The BIC outperforms the AIC and AICc and works down to a diffusion coefficient of $0.01\,\mu m^{2}/s$. (C) Percentage of correct decisions versus acquisition time (Parameters: $D_{input}=0.1\,\mu m^{2}/s$, $N=500\,$points, $B_{r}=30\,nm$, $k=0.3\,pN/\mu m$). The BIC outperforms the AIC and AICc and works for acquisition times between $1\,$ms and $1000\,$ms. (D) Percentage of correct decisions versus input spring constant (Parameters: $D_{input}=0.1\,\mu m^{2}/s$, $N=500\,$points, $B_{r}=30\,nm$). The BIC outperforms the AIC and AICc.  
}
\label{Fig3}
\end{figure}

This section studies the performance of the three information theory criteria for numerical Brownian trajectories that are confined in a harmonic-spring potential ($V=1/2kr^{2}$), where \textit{k} is the spring constant. Again, we simulated $300$ trajectories under each condition and recorded the normalized histogram of decisions, which are made for each trajectory by determining which model yielded the minimal value for a criterion. The performance of the BIC, AIC and AICc with respect to the trajectory length is shown in figure \ref{Fig3} (A). The BIC slightly outperforms both the AIC and AICc for most trajectory lengths and correctly attributes the right potential down to $50$ trajectory points. The diffusion coefficient does not change the performance of the BIC over a large range of D (Fig. \ref{Fig3} (B)). The AIC and AICc only perform down to a D of $0.05\,\mu m^{2}/s$. Figure \ref{Fig3} (C) shows that the BIC outperforms both the AIC and AICc and is correct over a larger range of acquisition times. The impact of the strength of the potential is investigated in Fig. \ref{Fig3} (D). The BIC is the better indicator and the performance remains constant, except for low \textit{k}. This investigation shows that the BIC performs overall better in determining the preferred model for Brownian trajectories that are confined by a spring potential. A histogram of decisions for parameters close to experimental conditions is given in Fig. S2.\\

\subsection*{Confined motion in an anharmonic potential ($V\propto r^{4}$)}

\begin{figure}[!ht]
\begin{center}
\includegraphics[width=4.86in]{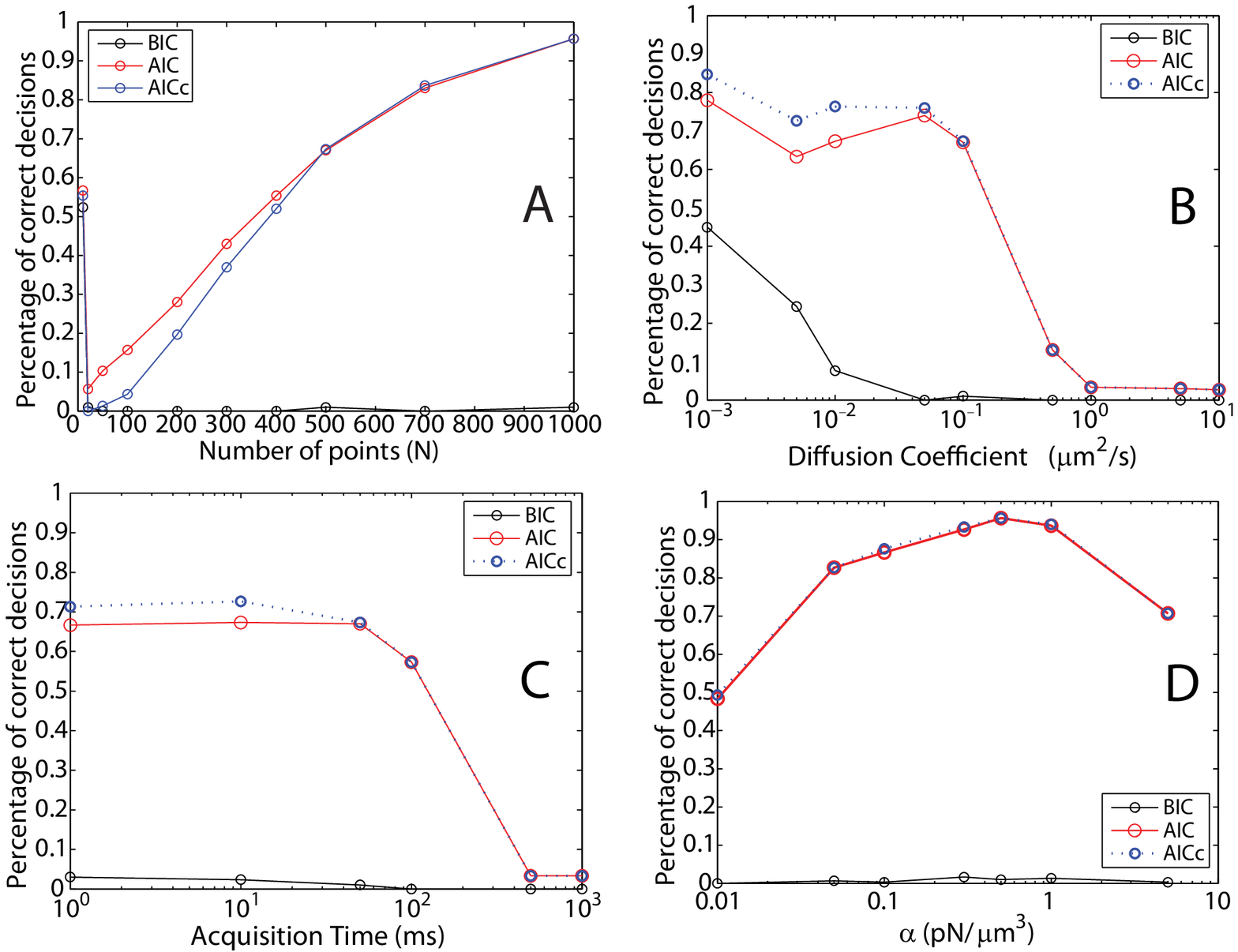}
\end{center}
\caption {
{\bf Information criteria for simulated Brownian trajectories confined in a $4$th order potential ($V=\alpha r^{4}$).} To determine the performance of the decision criteria, we calculated the BIC (black), AIC (blue) and AICc (red) for trajectories under various conditions. (A) Percentage of correct decisions ($300\,$trajectories per point) versus the length of the trajectory (Parameters: $D_{input}=0.1\,\mu m^{2}/s$, $t_{acq}=50\,ms$, $B_{r}=30\,nm$, $\alpha=0.5\,pN/\mu m^{3}$). The AIC slightly outperforms AICc for shorter trajectories. The BIC classifies the trajectory erroneously as confined in a $2$nd order potential. (B) Percentage of correct decisions versus the input diffusion coefficient (Parameters: $N=500\,$points, $t_{acq}=50\,ms$, $B_{r}=30\,nm$, $\alpha=0.5\,pN/\mu m^{3}$). The AICc slightly outperforms AIC for small diffusion coefficients. The BIC is wrong and classifies the trajectory erroneously as confined in a $2$nd order potential. The AIC and AICc cease to work for diffusion coefficients higher than $0.2\,\mu m^{2}/s$. (C) Percentage of correct decisions versus acquisition time (Parameters: $D_{input}=0.1\,\mu m^{2}/s$, $N=500\,$points, $B_{r}=30\,nm$, $\alpha=0.5\,pN/\mu m^{3}$). The AICc slightly outperforms AIC for short acquisition times. The BIC does not work, i.e. correct attribution below $50\%$. The AIC and AICc cease to work for acquisition times longer than $100\,ms$. (D) Percentage of correct decisions versus input potential strength (Parameters: $D_{input}=0.1\,\mu m^{2}/s$, $N=500\,$points, $B_{r}=30\,nm$). The AIC and AICc work equally well and outperform the BIC. Very weak potentials cannot be detected.   
}
\label{Fig4}
\end{figure}

The last investigated case are numerical Brownian trajectories that are confined by a 4th order (anharmonic) potential $V=\alpha r^{4}$. As before, we simulated $300$ trajectories under each condition and recorded the normalized histogram of decisions made for each trajectory. The performance of the BIC, AIC and AICc with respect to the trajectory length is shown in figure \ref{Fig4} (A). In stark contrast to the previous two cases, the BIC does not perform well and fails to attribute the correct potential. However, both the AIC and AICc do find the correct potential down to a trajectory length of $400$ points. As shown in Fig. \ref{Fig4} (B), the AIC and AICc can determine the correct model up to a diffusion coefficient of $0.1\,\mu m^{2}/s$, while the BIC cannot. The AIC and AICc can determine the correct model for acquisition times below $200\,ms$ when \textit{D} is $0.1\,\mu m^{2}/s$ (Fig. \ref{Fig4} (C)). Figure \ref{Fig4} (D) shows that, similarly to the spring-potential case above, the strength of the confining potential does not significantly influence the ability to choose the correct potential for the AIC and AICc, except at very low $\alpha$ values. These data suggest that the BIC cannot be used to determine that a trajectory is confined by a $4$th order potential. However, the AIC and AICc are good tools to do so.\\

\subsection*{Building the decision tree}

Because the BIC cannot always determine the correct model, as shown by the results above, model selection cannot be performed in a unique step with one criterion. As we can see, the BIC can distinguish between the free Brownian case and the harmonic potential. However, AIC or AICc are able to distinguish between the harmonic potential and the anharmonic potential over a wide parameter range. This information points towards building a decision tree with nested models that first distinguishes between free motion and confinement and then, in a second step, can determine the shape of the potential. Using the idea of the decision tree, we mapped a greater parameter space than previously explored to determine the structure and limits of the final decision tree. Simulations mapping the trajectory length \textit{N} and the diffusion coefficient \textit{D} space, are shown in figure \ref{Fig5}. Here, the normalized performance of the three criteria for $300$ numerical trajectories per condition is shown by a color code after a threshold of $0.5$ has been applied. The threshold is used because we require that our method should determine the correct model over half the time. The three models are shown vertically in columns and the three criteria BIC, AIC and AICc are shown for the three models in rows, as indicated. The bright areas show that the model could be attributed correctly and black squares indicate that the correct model could not be found, i.e. more than $50\%$ of trajectories were falsely classified.\\

As previously suspected, the BIC can correctly distinguish between free Brownian motion and the harmonic potential (red box), but fails to determine the anharmonic 4th order potentials. The AIC and AICc on the other hand, can distinguish between the different potentials (blue box), but cannot be used to determine free Brownian motion. Similar results are obtained by looking at the other parameters, such as acquisition time $t_{acq}$ and \textit{N} (Supporting Figure S3). The potential strength also does not change the fact that only the AIC and AICc can be used to distinguish the different potentials (Supporting Figure S$4$).\\

The key in building the decision tree was to realize that the anharmonic $4$th order potential was falsely determined to be a spring potential by the BIC, as shown in in Supporting figure S$5$. Thus, we used the BIC only to determine if a trajectory was undergoing free Brownian motion or if it was confined in a potential (red arm in Fig. \ref{Fig1} (bottom)). The simulations showed that this was possible for trajectory lengths down to $20$ points for most \textit{D} and acquisition times between $10$ and $200\,ms$. Once, a trajectory has been determined to be confined in a potential, the AIC or AICc can be used to classify the potential to be spring-like or $4$th order. For this classification, the trajectory length has to exceed $500$ points, the acquisition time has to be below $200\,ms$ and the potential strength of the $4$th order potential should be in the range of $0.1$ to $1\,pN/\mu m^{3}$. These conditions are the consequence of the information accessible in the trajectory as demonstrated in \cite{Voisinne2010}. If the tracked biomolecule moves too fast with respect to the acquisition time or if the force due to the potential is far greater than the thermal noise, it will become increasingly difficult to resolve these effects. Overall, the AIC gives a slightly larger window in which it determines the correct potential. Thus, we use the AIC instead of the AICc in our final decision tree. A source code that calculates the BIC, AIC and AICc for a given trajectory for the three models is provided in the supporting information (Model distinction in trajectory.c) (Source Code S11).\\

\subsection*{Application to experimental \textit{Clostridium Perfingens} $\epsilon$-toxin (CP$\epsilon$T) receptor trajectories}

We applied the derived decision tree to experimental \textit{Clostridium Perfingens} $\epsilon$-toxin receptor trajectories. This pore-forming toxin binds to a receptor in the cell membrane and undergoes confined motion in lipid rafts \cite{turkcan2012exp,masson2009inferring,turkcan2013flow}. The tracked toxin monomers exploit these confinement zones to locally increase their concentration and initiate further biological functions. In previous work, we have analyzed the trajectories by Bayesian inference and modeled them by a $2$nd or $4$th order potential. We then used a comparison of the magnitude of terms in the two polynomials to conclude that a $2$nd order description of the confining potential is sufficient because the $4$th order terms are small \cite{turkcan2012exp}. With our here improved decision tree technique, we analyzed a total of $60$ trajectories, which we have cut to only contain $500$ points. The first step of the decision tree that uses the BIC determined that $59$ trajectories are confined, while only $1$ trajectory was attributed to free Brownian motion (Fig. \ref{Fig6} insert). For the confined trajectories, the next step in the decision tree uses the AIC to determine the shape of the potential. The blue histogram in figure \ref{Fig6} shows that the majority of the experimental trajectories ($54$) were found to undergo confined motion in a spring potential, while only $5$ were found to be confined by a $4$th order potential.\\

The results of the decision tree agree with the previous findings that the CP$\epsilon$T receptor undergoes confined motion in a spring-like potential \cite{turkcan2012exp}. However, the present method is much more flexible and fast in classifying the shape of the potential than the previously used method which consisted in fitting the confining potential with polynomials of growing order and quantifying the evolution of the likelihood along with the error on the inferred parameters.\\

\subsection*{Preferred model selection with temporal resolution for multi-mode trajectories}

Single-molecule trajectories often change their mode of motion. Here, we applied the decision tree method to classify the mode of motion along a single-molecule trajectory to determine the current mode of motion during the observed trajectory.\\

\begin{figure}[!ht]
\begin{center}
\includegraphics[width=6.25in]{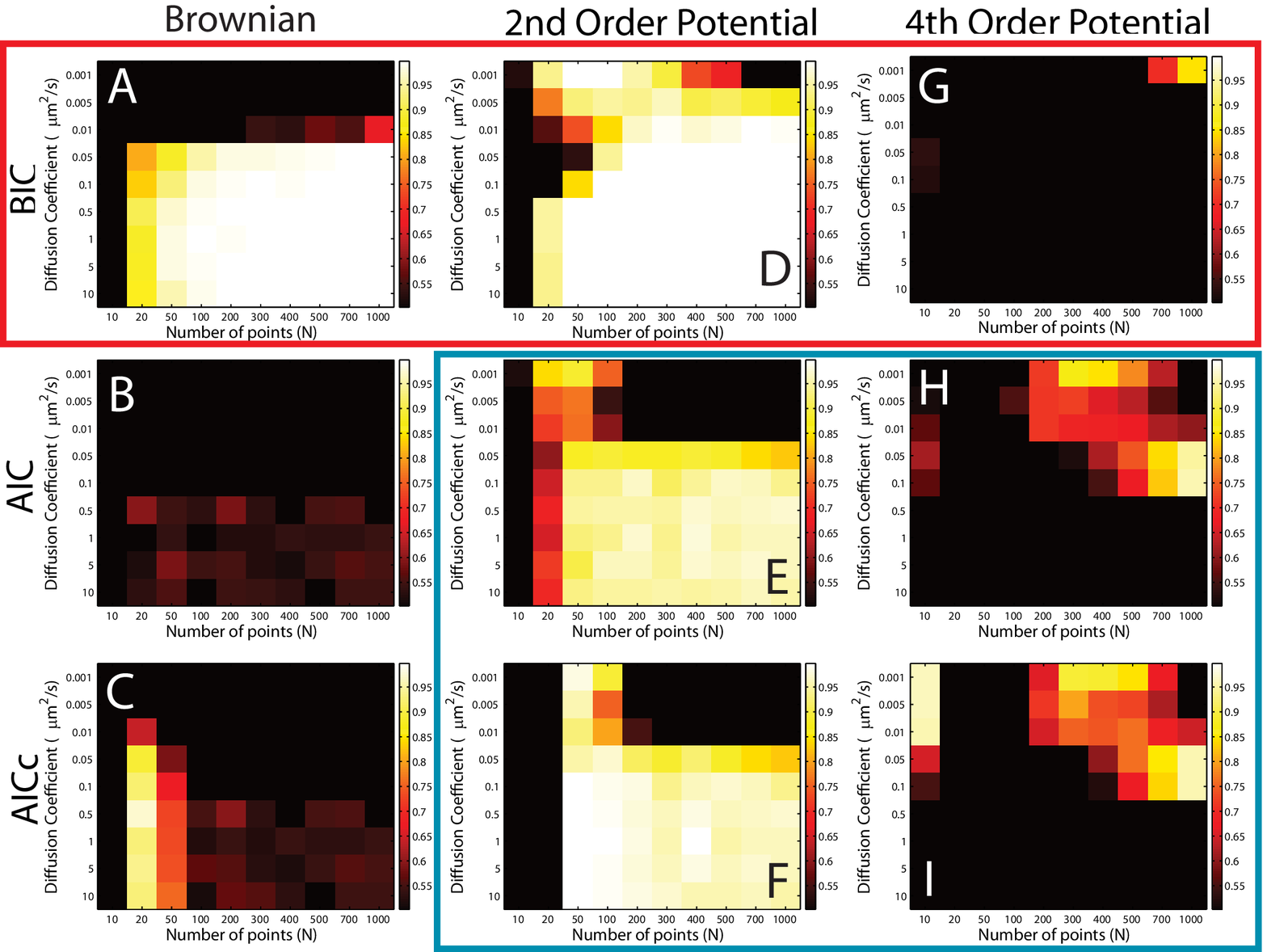}
\end{center}
\caption{
{\bf Building the decision tree using information criteria from simulated trajectories.} The $2$D plots show the heat map of the percentage of correct decisions out of $300$ simulated trajectories per square for the BIC (first row), AIC (middle row), and AICc (bottom row). The input trajectories were free Brownian (left column), Brownian confined in a $2$nd order spring potential $V\propto r^{2}$ (middle column), and Brownian confined in a $4$th order potential $V\propto r^{4}$ (right column). The heat map has a threshold of $0.5$, which means that only cases where the information criterion works correctly more than half of the time are non-black as indicated by the color scale. The BIC is the better criterion to determine if a trajectory is undergoing purely Brownian motion or if is confined by a potential (red box \& red arm in decision tree in Fig. \ref{Fig1}). The BIC is not suited to distinguish between a $2$nd and $4$th order potential. Here, the AIC and AICc provide a solution (blue box \& blue arm in decision tree in Fig. \ref{Fig1}). 
}
\label{Fig5}
\end{figure}

\begin{figure}[!ht]
\begin{center}
\includegraphics[width=3.27in]{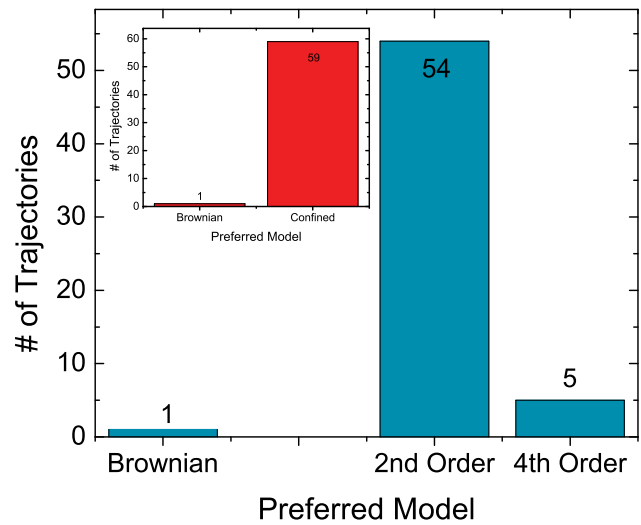}
\end{center}
\caption {
{\bf Classification of experimental Clostridium Perfingens $\epsilon$-toxin (CP$\epsilon$T) receptor trajectories.} We apply the decision tree in Fig. \ref{Fig1} to $60$ experimental trajectories with a length of $500$ frames. First we use the BIC to determine that the trajectories are confined (red insert). $59$ trajectories were found to be confined while one trajectory was attributed to free Brownian motion. The AICc shows that the  CP$\epsilon$T receptors are confined in a $2$nd order potential $V\propto r^{2}$, which is in agreement with previous results \cite{turkcan2012exp}.
}
\label{Fig6}
\end{figure}

\begin{figure}[!ht]
\begin{center}
\includegraphics[width=6.25in]{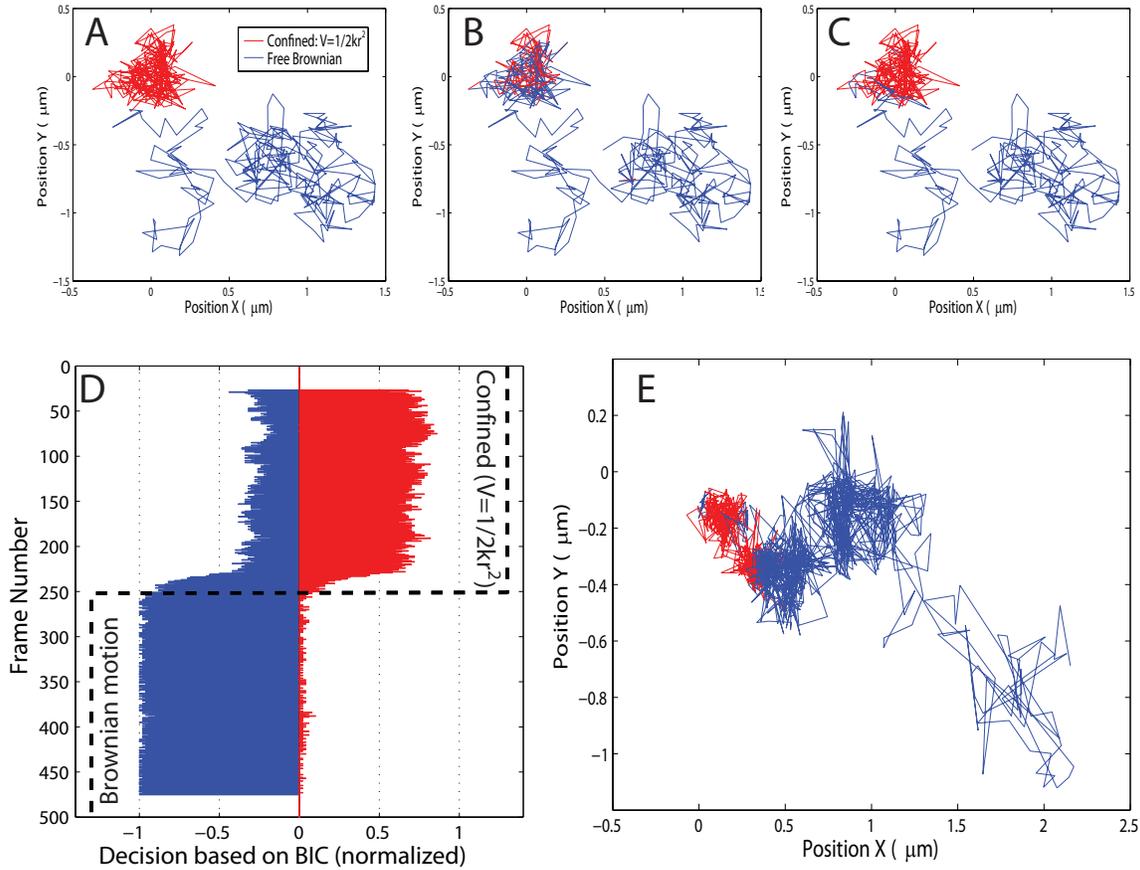}
\end{center}
\caption {
{\bf Classification of the mode of motion along a trajectory.} (A) We apply the first part of the decision tree in Fig. \ref{Fig1} to single numerical trajectories, which switch from being confined by a spring potential (red) to free Brownian motion (blue). (Parameters: $D_{input}=0.1\,\mu m^{2}/s$, $t_{acq}=50\,ms$, $B_{r}=30\,nm$, $k=0.3\,pN/\mu m$). (B) shows the result of using the BIC criterion along the numerical trajectory shown in (A). We use a window of $51$ frames that slides along the trajectory, and a classification is made for each central frame of the window. The method can correctly identify confinement (red). (C) Low-pass filtering the classifications gives a very robust method to determine the mode of motion of a trajectory that changes. (D) shows the performance of the BIC along the $500$ frames of $50$ numeric trajectories. The input mode is shown by the black dotted line, which is at first confined and switches to free Brownian motion at frame $250$. The blue histogram shows the number of free Brownian classifications at a certain central frame. The red histogram shows the number of spring-potential confined classifications. (E) Classification along a CP$\epsilon$T receptor trajectory, while confinement is reduced due to a modification of the cell membrane.
}
\label{Fig7}
\end{figure}

To this end, we selected the preferred model via the information criteria along a trajectory using a window of variable size. We chose simulation parameters, similar to experimental conditions ($D_{input}=0.1\,\mu m^{2}/s$, $t_{acq}=50\,ms$, $B_{r}=30\,nm$, $k=0.3\,pN/\mu m$, $N=500\,$frames). Consulting Fig. \ref{Fig5}, we chose a window size of $51$ frames that slides along the trajectory. As an example, we studied the transition from confined motion in a harmonic spring potential to free Brownian motion half way through the numerical trajectory (Fig. \ref{Fig7} (A)).\\

Figure \ref{Fig7} (B) shows the determined mode of motion for the numerical trajectory shown in \ref{Fig7} (A). The Bayesian decision tree can correctly identify the confined part (red) of the trajectory and the free Brownian motion part (blue). A low-pass filter that does not allow switching of modes unless three consecutive frame positions of the window yield a mode change, gives a very reliable result (Fig. \ref{Fig7} (C)).\\

To study the performance in greater detail, we evaluated the mode of motion of $50$ numerical trajectories. Figure \ref{Fig7} (D) shows the number of decisions, using the BIC criterion (shown on the x-axis as a normalized histogram), as the central frame of the 51-frame time window slides along the trajectories (shown on the y-axis). When a frame was deemed free Brownian, it is represented in blue (left side), when it was deemed confined in a spring potential it is represented in red (right side). The red and blue counts add up to unity for each central frame. The input mode of motion is shown by the black dotted line and switches from confined to free at frame $250$. On average, the method can correctly identify the mode of motion. The BIC is more sensitive towards free Brownian motion, i.e. when some part of the $51$-frame long window of observation sees free Brownian motion, the decision is deemed Brownian overall. This explains why the method finds the frame at which the trajectory switches from confined to Brownian at frame number $228\pm 9$ for the ensemble of $50$ trajectories. The ratio of confined to free Brownian motion decisions agrees with the ratio previously determined by the simulations for the decision tree (Fig. \ref{Fig5}).\\

\subsection*{Comparison of the Bayesian decision tree method to a residence time based method}

\begin{figure}[!ht]
\begin{center}
\includegraphics[width=6.25in]{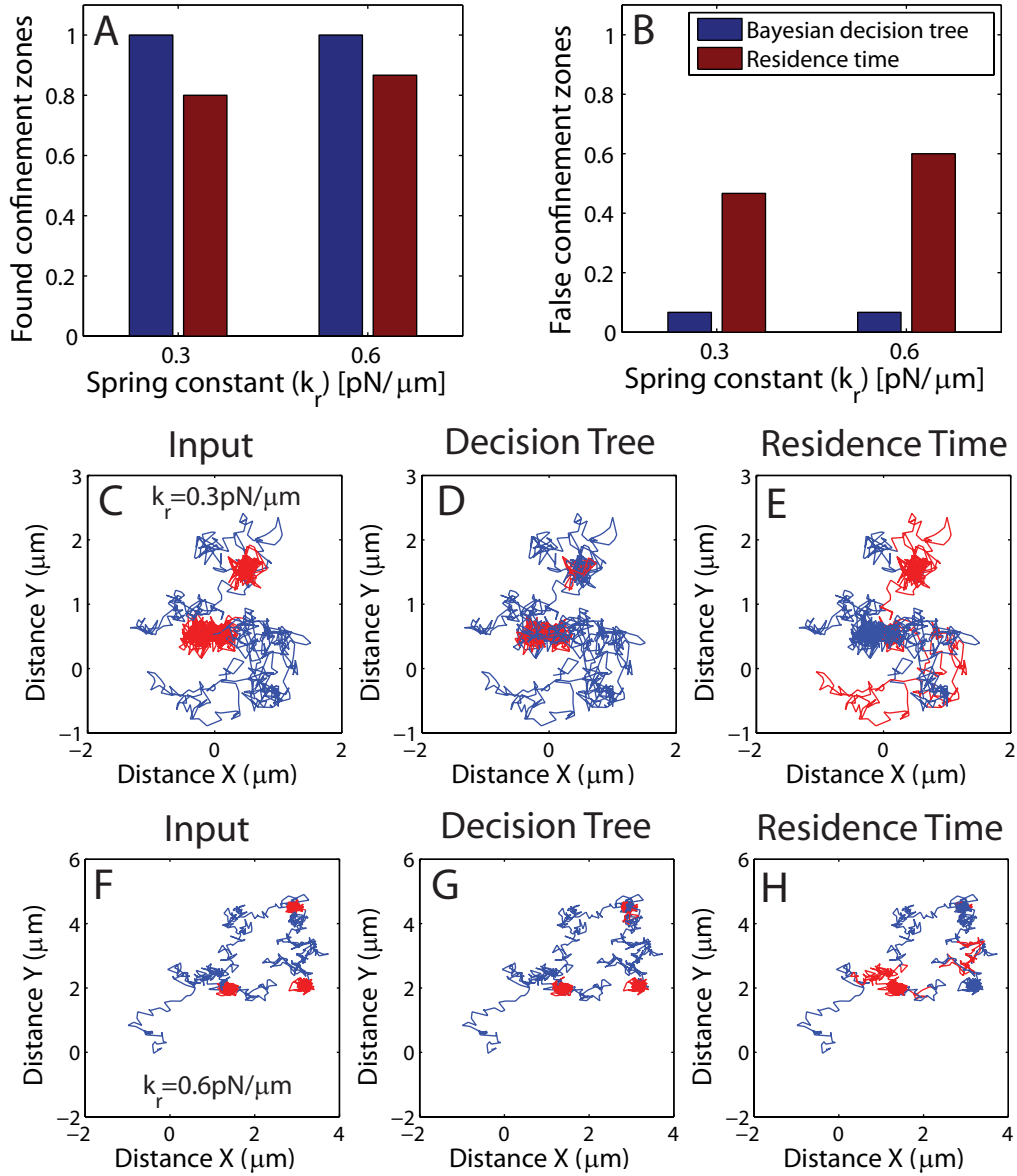}
\end{center}
\caption {
{\bf Comparison to the residence time method.} We apply the first part of the decision tree in Fig. \ref{Fig1} to single numerical trajectories, which cycle between confinement by a spring potential and free Brownian motion. (Parameters: $D_{input}=0.1\,\mu m^{2}/s$, $t_{acq}=50\,ms$, $B_{r}=30\,nm$, $k=0.3$ \& $0.6\,pN/\mu m$). Receptors are confined three times for 200 frames (10s). (A) Histogram of the correctly found confinement zones out of 15 zones for the Bayesian decision tree method (blue) and the residence time method (red) for two spring constants. We use a window of $51$ frames that slides along the trajectory, and a classification is made for each central frame of the window. (B) shows a histogram of non-existing found confinement zones for the Bayesian decision tree method (blue) and the residence time method (red) for two spring constants. (C \& F) One of the five input trajectories with three confining zones (red) with a spring constant of $k=0.3\,pN/\mu m$ and $k=0.6\,pN/\mu m$, respectively. (D \& G) Result using the Bayesian decision tree method with free brownian motion in blue and confined motion in red. (E \& H) Result using the residence time method with free brownian motion in blue and confined motion in red.}
\label{Fig8}
\end{figure}

To further evaluate the precision of the here presented decision tree method we compare it to a commonly used residence time based method. The residence time based method detects temporal confinement by identifying periods in which the receptor remains in a membrane region for durations longer than a free Brownian diffusing particle would stay in an equally sized region \cite{simson1995detection}. We generate $5$ trajectories for two different spring constants of $1200$ frames length ($60\,s$) that have three distinct temporal confinement zones of $200$ frames duration ($10\,s$) and test whether the two methods can detect the real confinement zones (Fig. \ref{Fig8} (A)) and the level of falsely detected confinement zones, which are not present in the input (Fig. \ref{Fig8} (B)). The results show that both methods can identify confinement zones but only the Bayesian decision tree method can detect all of the $15$ input regions. The real strength of the Bayesian decision tree method lies in the low number of false positives, when compared to the residence time based method. The decision tree method only finds one non-existing confinement region, while the residence time method finds $5$ and $6$ false zones for a spring constant of $0.3$ and $0.6\,pN/\mu m$, respectively. Figure \ref{Fig8} (C-E) and (F-H) show the performance of the two methods along two numeric sample trajectories for two different confining spring constants of $0.3$ and $0.6\,pN/\mu m$, respectively. The residence time based method in Fig. \ref{Fig8} (E \& F) fails to detect one input zone and falsely declare a free Brownian section of the trajectory as confined.\\

\subsection*{Model selection along experimental Clostridium Perfingens $\epsilon$-toxin (CP$\epsilon$T) receptor trajectories}

We applied the preferred model selection with temporal resolution to an experimental CP$\epsilon$T receptor trajectory that changes its mode of motion. As previously shown, the confining membrane domain of the CP$\epsilon$T receptor can be destabilized by oxidizing cholesterol in the cell membrane \cite{turkcan2013probing}.\\

The Bayesian decision tree method determines a change of motion from confined to free Brownian motion along the recorded experimental data (Fig. \ref{Fig7} (E)). Here, the trajectory is undergoing confined motion (red) at the beginning of the incubation with the enzyme cholesterol oxidase. An additional trajectory is shown in the Supporting Figure S6. As the enzyme gradually oxidizes more cholesterol in the membrane, the receptor switches to free Brownian motion (blue). This result agrees with our previous work, which detected a decrease in confining potential along the CP$\epsilon$T receptor trajectory during incubation with cholesterol oxidase.\\

\section*{Discussion}

We introduced a decision-tree based method that uses information criteria to determine the mode of motion of a single-molecule trajectory. The method is based on the combined use of the Bayesian information criterion (BIC), the Aikaike information criterion (AIC), and a modified version of the AIC (AICc). These criteria are used to determine which model best describes a specific data set. The models that we discussed in this work are free Brownian motion, confined motion in a harmonic potential and confined motion in a $4$th order polynomial potential. All of these models can be associated to various identifiable structural characteristics. A spring-like potential can indicate actin tethering or hydrophobic interactions and an anharmonic $4$th order potential can indicate more local confinement by the picket-fence model, for example. The information criteria allows association of experimental data to these models. Thus, this method can provide quantitative information on how much better a certain model describes a trajectory with respect to other competing models. Of course, it is possible to extend this method to include more models of motion. Furthermore, more complex environments can be numerically investigated and associated to more simplified models, such as the ones discussed here to classify trajectories. \\

In order to build the decision tree, numerical trajectories were generated under various conditions matching previous experimental data in the field of single-molecule tracking. These trajectories were either undergoing free Brownian motion or confined motion in a potential. For each trajectory, we calculate the BIC, AIC and AICc using the posteriori probabilities corresponding to each investigated model. Then the algorithm chose the preferred model by evaluating which model has the smallest information criterion value. The key feature was that this method supplied a reliable method based on simple criteria that can automatically classify the mode of motion.\\

Surprisingly, the most accurate method in choosing the preferred model is not a simple decision making step, but a decision tree where the two confined models are nested. A first decision using the BIC determines if the trajectory is undergoing free Brownian motion or if it is confined. Here, the minimal trajectory length for meaningful classification is $50$ frames under most conditions (D: $0.01-10\,\mu m^{2}/s$, k$<0.1\,pN/\mu m$, and $t_{acq}$: $10-100\,ms$). In a second step, the nature of the confining potential is investigated. The AIC is used to distinguish between the harmonic potential and the anharmonic potential. This distinction can be made with trajectories that have more than $500$ frames for diffusion coefficients below $0.1\,\mu m^{2}/s$, spring constants of $0.02-10\,pN/\mu m$, and $\alpha$ in the range of $0.1-1\,pN/\mu m^{3}$. Experimentally determined single molecule diffusion coefficients range from $0.004$ to $2.2\,\mu m^{2}/s$ \cite{sheets1997transient,ritchie2005detection}. Thus, the presented method yield reliable results for two out of the three magnitudes spanned by results from literature. Experimental spring constants lie in the range of $0.3-10\,pN/\mu m$ \cite{Jin2007,oddershede2002,Peters1999,turkcan2012exp}. The method can provide an accurate model distinction for the entire range of experimental spring constant values in the current literature. The need for at least $500$ frames is the largest limitation of the presented technique and will provide a challenge for groups that use organic fluorophores for tracking. However, quantum dots or fluorescent nanoparticles provide an alternative label that is more photostable. It should be noted that we did not discuss the accuracy of the inferred parameters, such as the diffusion coefficient in this text and we only focus on the ability to attribute the correct model. They might be subject to a bias, as discussed in reference \cite{turkcan2012bayesian}. We then apply the decision tree to experimental CP$\epsilon$T receptor trajectories and show that the method can reliably confirm previous findings about the nature of the confining potential of these receptors.\\

The method can also be adapted to determine changes in the mode of motion along a single-molecule trajectory. To this end, we evaluated the mode of motion using a window of $51$ frames that slides along the trajectory in time. Simulations show that the method can reliably classify confinement regions on numerical trajectories and that it outperforms the commonly used residence time based method. Thus, this technique supplies a rigorous and reliable tool to automatically segment and classify SMT data. Finally, this method can easily confirm previous findings about the loss of confinement of CP$\epsilon$T receptor trajectories due to lipid raft disaggregation. In the previous analysis, the mean diffusion coefficient increased from $0.063\pm 0.01\,\mu m^{2}/s$ to $0.18\pm 0.02\,\mu m^{2}/s$, while the spring constant of the confining potential decreased from $237\pm 44\,k_{B}T. \mu m^{-2}$ to $35.4\pm 7.7\,k_{B}T. \mu m^{-2}$ \cite{turkcan2013probing}. In the previous work, we have analyzed the change in motion of many receptors and concluded that their confinement reduces but does not become entirely free Brownian motion \cite{turkcan2013probing}. Here, we have selected the most extreme trajectories from the dataset, which do switch to free Brownian motion for the sake of showing a change in the mode of motion.\\

In conclusion, we have successfully built a method that can reliably distinguish between different modes of motion over a wide parameter range. Furthermore, it is possible to quantify how reliable the method is for each group of input parameters. This method can be extended to include further models of motion, for example phenomena like active transport and hopping.\\

This technique adds another dimension to the discussion of SMT data. Currently, most of the discussions are focused on determining the mean value of quantities, such as diffusion coefficients and confining domain sizes. Such measurements, however, are not easily compared over the vast range of experiments and tracked biomolecule species. Additionally to parameters, such as diffusion coefficient and domain sizes, this method can be used to gain information about the mode of motion and changes in the mode of motion. The mode of motion of a receptor might hold more biologically relevant information than the diffusion coefficient or domain size and is perhaps a better tool to classify and compare different SMT experiments.\\

\section*{Materials and Methods}

\subsection*{Generating numerical trajectories}

In this work, we generate numerical trajectories that undergo free Brownian motion or Brownian motion in a 2nd and 4th order potential ($V\propto r^{2}$ \& $V\propto r^{4}$). To simulate two-dimensional Brownian motion, the length of each step was taken from a Gaussian distribution with a standard deviation of $\sqrt{4D\Delta t}$, where the input diffusion coefficient $D$ and the acquisition time $\Delta t$. The angle of each step is randomly distributed over $[0,2\pi]$. Each particle takes $1000$ substeps during each $\Delta t$. The substeps are not averaged. 
If the trajectory is confined by a potential, the displacement due to the force generated by the confining potential is added to each substep. The confining potentials used, as demonstration, in this work are the 2nd order spring potential ($V(x,y)=1/2k_{X}x^2+1/2k_{Y}y^2$ with $k=\sqrt{k^{2}_{X}+k^{2}_{Y}}$), and the 4th order potential ($V(x,y)=\alpha_{X}x^4+\alpha_{Y}y^4$ with $\alpha=\sqrt{\alpha^{2}_{X}+\alpha^{2}_{Y}}$). Yet, we emphasize that this approach is not limited to spring potentials nor to polynomial potentials. For all numerical trajectories, static positioning noise $B_{r}=30\,nm$ was added to the trajectory by an additional displacement taken from a Gaussian distribution with standard deviation $2B_{r}$ with an angle randomly distributed over $[0,2\pi]$. This Gaussian noise models all sources of noise, i.e. Poissonian photon shot noise due to signal and fluorescence background, detector noise, pixelization effects, and error of the localization algorithm using a Gaussian representation. The source code for the trajectory generation is given in the supporting information: Free Brownian motion (GenerateBrownianTraj.c) (Source Code S8), Brownian motion confined in a 2nd order spring potential (GenerateBrownianTrajin2ndOrderPot.c) (Source Code S9), and Brownian motion confined in a 4th order potential (GenerateBrownianTrajin4thOrderPot.c) (Source Code S10).

\subsection*{Bayesian Inference}

We developed the Bayesian inference approach to treat single-molecule trajectories in previous works \cite{masson2009inferring,turkcan2012bayesian}, but we include some information here to make this article self-contained.\\

The single-molecule motion is modeled by the overdamped Langevin equation:\\

\begin{eqnarray}
\frac{d\textbf{r}}{dt}=-\frac{\nabla V(\textbf{r})}{\gamma(\textbf{r})}+\sqrt{2D(\textbf{r})}\xi(t),
\label{langevin}
\end{eqnarray}
\\

with $\gamma (\textbf{r})$ the spatially varying friction coefficient, $D(\textbf{r})$ the spatially varying diffusion coefficient, $V(\textbf{r})$ the potential acting on the biomolecule and  $\xi(t)$ the rapidly varying zero-average Gaussian noise. The fluctuation-dissipation theorem gives $D(\textbf{r})=k_{B}T/\gamma(\textbf{r})$ \cite{risken1996fokker}. In this work we don't address the question of diffusion heterogeneities by setting $D(\textbf{r})=D$.\\

The associated Fokker-Planck equation, which governs the evolution of the transition probability over time, has no general solution for an arbitrary potential and a spatially varying diffusion coefficient. We therefore divide the confinement domain into subdomains using a mesh grid and the points of the trajectory are attributed to their respective grid subdomains. Within each subdomain, we consider that the potential gradient is constant. Note, that this mesh is not used, when the suspected model is a purely Brownian trajectory. This assumption enables us to solve the associated Fokker-Planck equation to Eq. \ref{langevin}, for a constant $F_{ij}$ and \textit{D} per subdomain $(i,j)$, where the forces in adjacent subdomains is free to vary. The assumption leads to the expression of the transition probability,

\begin{eqnarray}
P_{Confined}((\textbf{r}_{2},t_{2}|\textbf{r}_{1},t_{1})|\mathbf{F}_{ij},D)=\frac{e^{-\frac{(\textbf{r}_{2}-\textbf{r}_{1}-\textbf{F}_{ij}(t_{2}-t_{1})/\gamma_{ij})^2}{4\left( D + \sigma^{2}/\left(t_{2}-t_{1}\right)\right)(t_{2}-t_{1})}  } }{4\pi \left( D + \sigma^{2}/\left(t_{2}-t_{1}\right)\right)(t_{2}-t_{1})}\label{Props}
\end{eqnarray}
\\

with $\sigma$ the amplitude of the positioning noise $2B_{r}$. This expression is the probability of going from one space-time coordinate (r$_{1}$, t$_{1}$) to the next (r$_{2}$, t$_{2}$) for a diffusivity  \textit{D} and a force $\textbf{F}_{ij}$ with a positioning noise $\sigma$.
The transition probability for the free Brownian motion can be obtained from Eq. \ref{Props} by setting $\mathbf{F}_{ij}=0$ and gives:

\begin{eqnarray}
P_{Brownian}((\textbf{r}_{2},t_{2}|\textbf{r}_{1},t_{1})|,D)=\frac{e^{-\frac{(\textbf{r}_{2}-\textbf{r}_{1})^2}{4\left( D + \sigma^{2}/\left(t_{2}-t_{1}\right)\right)(t_{2}-t_{1})}  } }{4\pi \left( D + \sigma^{2}/\left(t_{2}-t_{1}\right)\right)(t_{2}-t_{1})}\label{Prop}
\end{eqnarray}
\\

The overall probability of observing a certain trajectory for a given set of variables is then computed by multiplying all the displacement probabilities between all individual points in the dataset, assuming that the motion of the molecule is a Markov process.

Now that the likelihood is known, we may apply Bayes' rule:

\begin{eqnarray}
P(Q|T)=\frac{P(T|Q)P_{0}(Q)}{P_{0}(T)},
\label{BayesTheo}
\end{eqnarray}
\\

where $P\left(Q|T\right)$ is the posterior or {\it a posteriori} probability of the parameters, {\it i. e.} the probability that the parameters \textit{Q} take on a specific value given the recording of the trajectory \textit{T}. $P\left(T|Q\right)$ is the likelihood of the trajectory, {\it i. e.} the probability of recording the trajectory \textit{T} given a specific value \textit{Q} of the parameters, $P_{0}\left(Q\right)$ the prior probability of the parameters and $P_{0}\left(T\right)$ is a normalization constant called the evidence of the model \cite{McKay,Udo_von_Toussaint}. Without prior knowledge on the parameters, the prior probability $P_{0}\left(Q\right)$ is supposed to be constant over a broad range of possible values.\\

\subsection*{Calculation of the posteriori probability and inferred value}

The inferred values are obtained from the posteriori probability distribution. All variables are initialized at zero. The prior probability $P_{0}\left(Q\right)$ is taken to be unity. A quasi-Newtonian optimization using the Broyden-Fletcher-Goldfarb-Shanno (BFGS) algorithm \cite{press2001numerical} finds the maximum of the {\it a posteriori} distribution (MAP) $P\left(T|D,\mathbf{F}_{ij}\right)$ on a 10 x 10 mesh of subdomains. In the case of inferring the $2$nd and $4$th order potential, the $\mathbf{F}_{ij}$ are no longer independent but are governed by $2$nd and $4$th order polynomials, respectively. The number of free parameters for the pure Brownian case is thus $1$, for the $2$nd order potential it is $6$, for the $4$th order it is $15$. The inference algorithms were programmed in C language and executed on a local PC (dual-core 3 GHz, 2 GB RAM) or Amazon web services (AWS) on c$1$.medium instances (high-CPU computing instance).\\

\subsection*{Calculation of Information criteria along the trajectory}

We determine the mode of motion along a trajectory that changes its mode of motion by calculating the BIC, AIC and AICc for a window of $51$ frames, which slides along the trajectory in time. The information criteria are calculated for each central frame \textit{i} of the window, by evaluating the posteriori probability and subsequently the information criteria for the trajectory portion between frame $i-25$ to frame $i+25$. We only distinguish between free Brownian motion and confinement in a spring potential. To test the algorithm, we generate $50$ numerical trajectories with parameters: $D_{input}=0.1\,\mu m^{2}/s$, $t_{acq}=50\,ms$, $B_{r}=30\,nm$, $k=0.3\,pN/\mu m$, $N=500\,$frames. Fig. \ref{Fig7} shows that the decision tree can distinguish these two modes of motion under these conditions.

\subsection*{Clostridium Perfingens $\epsilon$-toxin (CP$\epsilon$T) receptor tracking}

Y$_{0.6}$Eu$_{0.4}$VO$_{4}$ nanoparticles were coupled to toxins as described in \cite{Casanova2007}. In brief, we coupled APTES-coated europium-doped nanoparticles to $\epsilon$-toxin produced by \textit{C. perfringens} bacteria (Cp$\epsilon$T) via the amine-reactive cross-linker bis (sulfosuccinimidyl) suberate (BS$_{3}$), as described in \cite{turkcan2012exp}. A BCA test used to determine the amount of toxin after the coupling process, showed a toxin:nanoparticle coupling ratio of 3:1.\\

Tracking experiments were performed with a wide-field inverted microscope (Zeiss Axiovert $100$) equipped with a $63$x, $NA = 1.4$ oil immersion objective and a EM-CCD (Roper Scientific QuantEM:$51$2SC). NPs were excited with an Ar$^{+}$-ion laser using the $465.8\,$nm line. A 500DRLP beam splitter  (Omega) directs the beam towards the sample. Emission was collected through the beam splitter and a $617/8$M filter (Chroma). Confluent cells on coverslips were incubated with $0.04\,nM$ NP-labeled proto-toxin (CP$\epsilon$pT) for $20$ minutes at  $37\,^{\circ}$C. The concentration was low to avoid oligomerization and observe single NPs ($\approx 10$ per cell). The sample was then rinsed three times with observation medium (HBSS + 10mM HEPES buffer, $1\%$ FCS) to remove non-bound toxins and nanoparticles. We recorded images at a frame rate of about $20\,Hz$ (exposure time: $50\,ms$; readout time: $1.3\,ms$) and an excitation intensity of $0.25\,kW/cm^{2}$ at $37\,^{\circ}C$ in observation medium. We have verified that the nanoparticles do not perturb the motion of the CP$\epsilon$T receptor through tests with substitute organic fluorophore labeling and the use antibodies \cite{turkcan2012exp}.\\

The toxin receptor position in each frame was determined from a Gaussian fit to the diffraction pattern of the nanoparticles with a home-made Matlab V$8.2$ (Mathworks, Natrick MA) algorithm.

\subsection*{MDCK Cell Culture}

Madin-Darby canine kidney (MDCK) cells were cultured in (DMEM, $10\%$ fetal calf serum (FCS), $1\%$ penicillin-streptomycin) culture medium at $37\,^{\circ}$C. For tracking experiments, cells were trypsinated two days before and transferred onto acid-bath treated glass microscope coverslips and grown until confluent. The medium was replaced by an observation medium (HBSS + $10mM$ HEPES buffer, $1\%$ FCS, pH 7.4) just before the tracking experiment, which lasted less than $1.5$ h.

\subsection*{Cell treatment with Cholesterol oxidase}
Where mentioned, we incubated cells with $20\,U/mL$ cholesterol oxidase (Calbiochem) in HBSS+$10 mM$ HEPES for $30$ minutes. A cholesterol quantification kit (Invitrogen) was used to determine successful cholesterol digestion on lyzed cells before and after incubation. The data has been previously used to examine membrane-protein interactions \cite{turkcan2013probing}.

\subsection*{Residence time method for the detection of temporary lateral confinement}
We followed the work by Simson et al. \cite{simson1995detection}. The probability $\Psi$ that a protein with diffusion coefficient \textit{D}, will stay in a circular region with radius \textit{R} for a time period $t_{wd}$  is calculated by:

\begin{eqnarray}
log\,\Psi=0.2048-2.5117\,Dt_{wd}/R^{2}
\label{logPSI}
\end{eqnarray}
\\

We compute this probability along the trajectory for a window of $51$ frames ($t_{wd}=2.55\,s$), with \textit{R} being the maximal displacement from the center position of the $51$ frame window and known input diffusion coefficient. To accentuate the region of nonrandom behavior, we calculate the probability level \textit{L} according to:

\begin{dmath}
L =
\left\lbrace
      \begin{array}{ll}
       -log(\Psi)-1  &\condition[]{for $\Psi \leq 0.1$},
        \\
       0             &\condition[]{for $\Psi > 0.1$},
      \end{array}
  \right.
\label{logPsI}
\end{dmath}

\textit{L} is filtered in magnitude with a cutoff $L_{c}=4$ and temporal $t_{c}=2.5\,s$. A plot of \textit{L} along a numerical sample trajectory is shown in Fig. S7. To qualify as a confinement zone, \textit{L} has to be larger than $L_{c}$ for a time greater than $t_{c}$. The parameters were optimized to detect the confinement zones without generating many false positives.

\section*{Acknowledgments}
We thank Dr. Antigoni Alexandrou for the critical review of this manuscript and for her helpful insight. We also thank Maximilian Richly for his support during the cholesterol oxidase treatment SMT experiments.

\bibliography{libraryfinal}

\section*{Figures}

\clearpage

\section*{Supporting Information}

\setcounter{figure}{0}
\makeatletter 
\renewcommand{\thefigure}{S\@arabic\c@figure}
\makeatother

\begin{figure}[!ht]
\begin{center}
\includegraphics[width=6.25in]{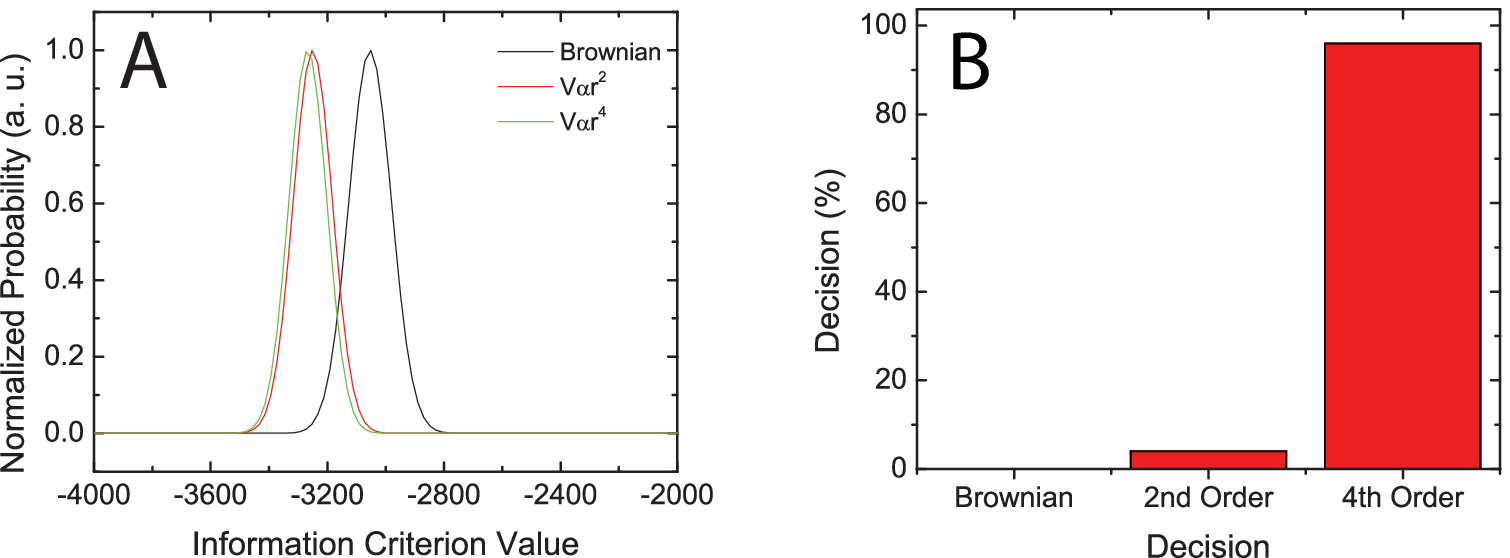}
\end{center}
\caption {
{\bf Averaged versus single-trajectory decision making.} (A) Calculating and averaging the information theory criteria, such as the here displayed AIC, for a distribution of 100 numerical  trajectories (Parameters: $D=0.1\,\mu m^{2}/s$, $N=500\,$points, $t_{acq}=50\,ms$, $B_{r}=30\,nm$, $\alpha=0.5\,pN/\mu m^{3}$) shows that it is possible to distinguish between free Brownian motion and confined motion, but it is impossible to determine the nature of the confining potential from the averaged data. (B) However, making decisions based on the criteria for each individual trajectory can lead to a histogram that correctly identifies the input potential.}
\label{FigS1}
\end{figure}

\begin{figure}[!ht]
\begin{center}
\includegraphics[width=3.25in]{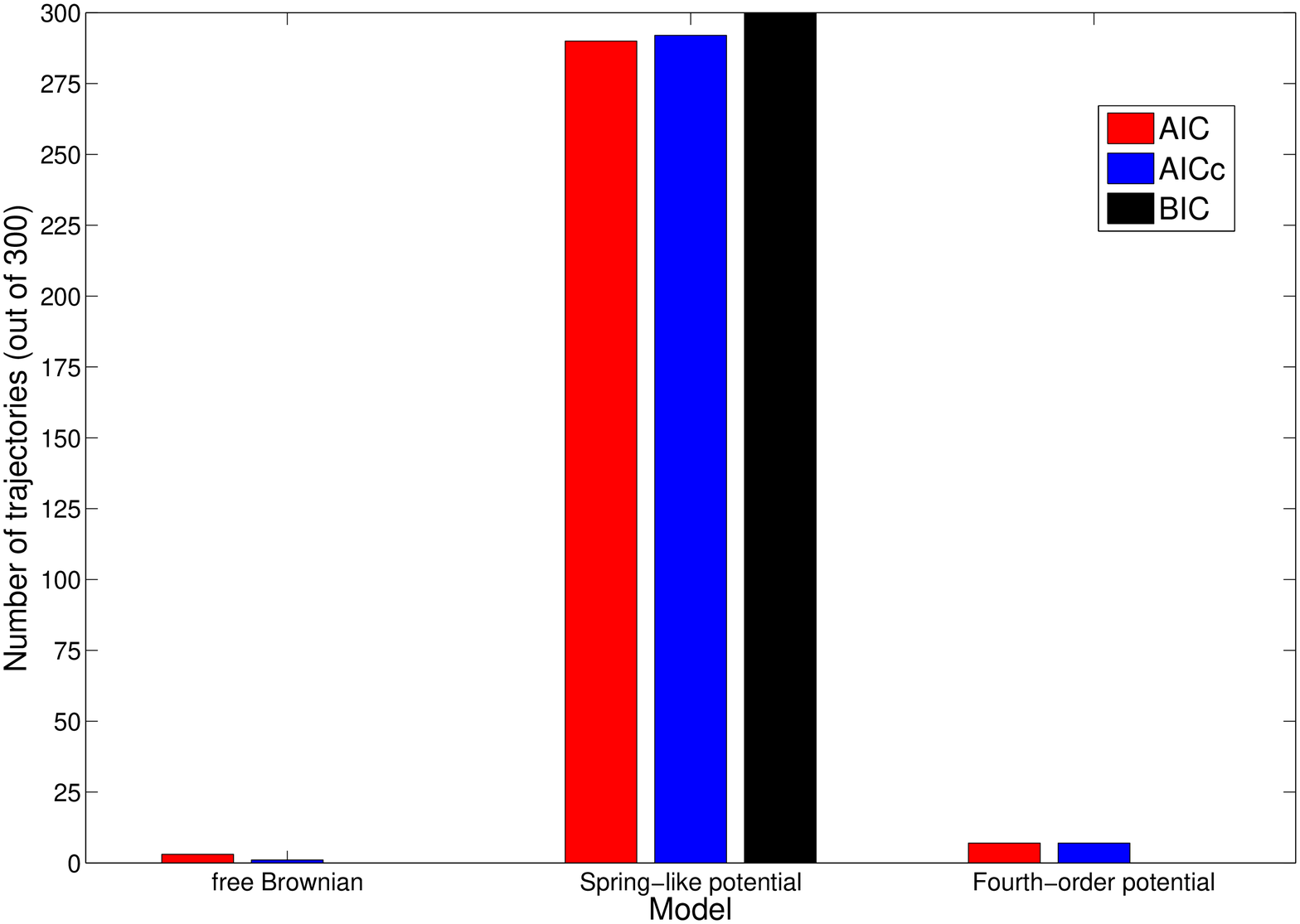}
\end{center}
\caption {
{\bf Histogram of trajectory classifications out of $300$ numerical input trajectories that resemble experimental trajectories with confinement in a spring-like potential.} (Parameters: $D=0.1\,\mu m^{2}/s$, $N=500\,$points, $t_{acq}=50\,ms$, $B_{r}=30\,nm$, $k=0.3\,pN/\mu m$). Decisions based on the BIC are shown in black, decisions based on the AIC and AICc are shown in red and blue, respectively.}
\label{FigS2}
\end{figure}

\begin{figure}[!ht]
\begin{center}
\includegraphics[width=6.25in]{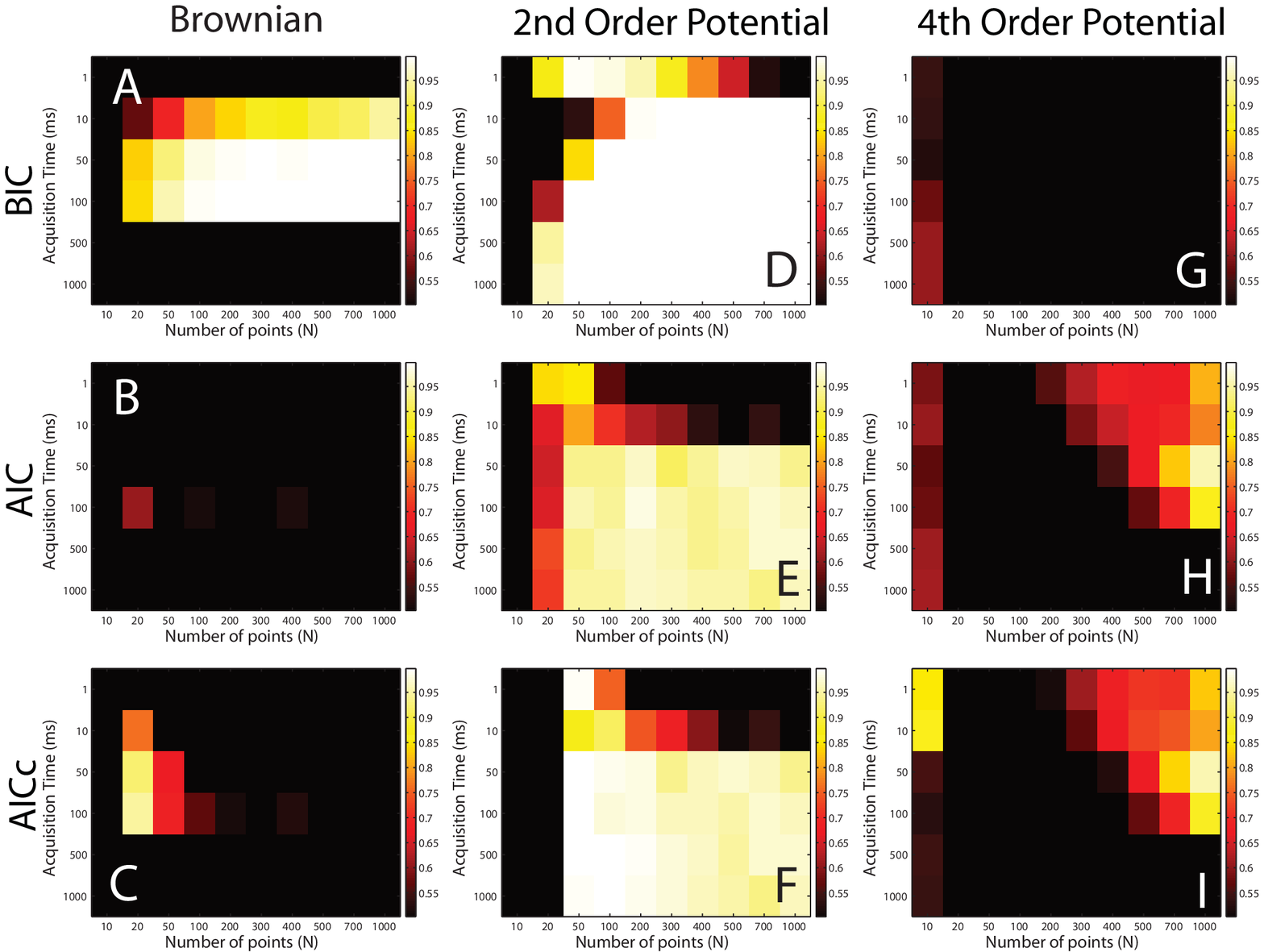}
\end{center}
\caption {
{\bf Building the decision tree using information criteria from simulated trajectories.}  The 2D plots show the heat map of the percentage of correct decisions out of $300$ simulated trajectories per square for the BIC (first row), AIC (middle row), and AICc (bottom row). The input trajectories were Brownian (left column), confined in a spring potential $V=1/2\,k r^{2}$ (middle column), and confined in a $4$th order potential $V=\alpha r^{4}$ (right column). The heat map has a threshold of $0.5$, which means that only cases where the information criterion works correctly more than half of the time are non-black as indicated by the color scale. The BIC is the better criterion to determine if a trajectory is undergoing purely Brownian motion or if is confined by a potential (red box \& red arm in decision tree in Fig. $1$). The BIC is not suited to distinguish between a $2$nd and $4$th order potential. Here, the AIC and AICc provide a solution (blue box \& blue arm in decision tree in Fig. $1$).}
\label{FigS3}
\end{figure}

\begin{figure}[!ht]
\begin{center}
\includegraphics[width=6.25in]{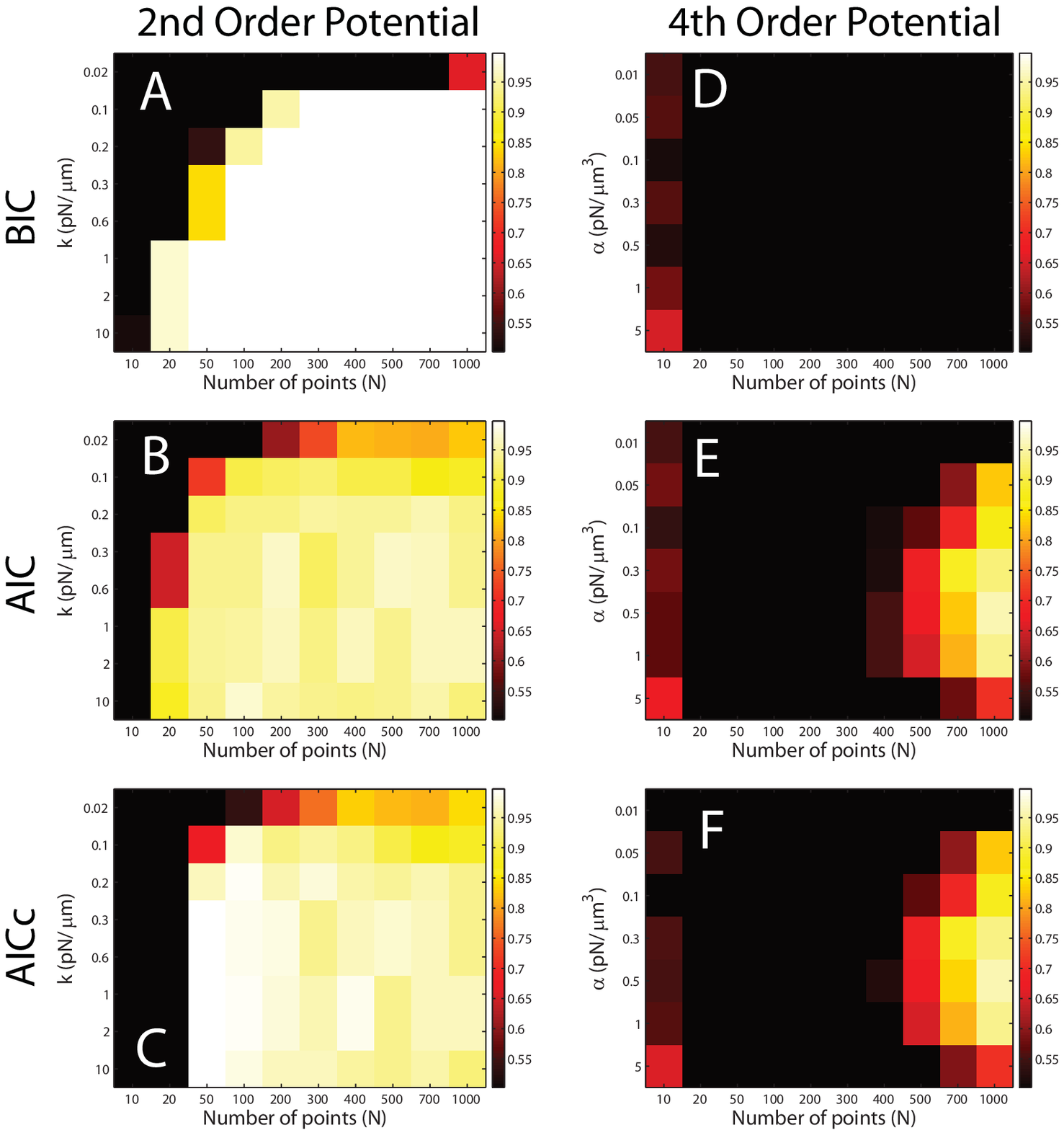}
\end{center}
\caption {
{\bf Building the decision tree using information criteria from simulated trajectories.} The 2D plots show the heat map of the percentage of correct decisions out of $300$ simulated trajectories per square for the BIC (first row), AIC (middle row), and AICc (bottom row). The input trajectories were Brownian walkers, confined in a spring potential $V=1/2\,k r^{2}$ (left column A-C), and confined in a $4$th order potential $V=\alpha r^{4}$ (right column D-F). The heat map has a threshold of $0.5$, which means that only cases where the information criterion works correctly more than half of the time are non-black as indicated by the color scale. The AIC and AICc are the only effective indicator that can distinguish between these two potential types. However, the strength of the potential does not have a large impact on the performance of these criteria.}
\label{FigS4}
\end{figure}

\begin{figure}[!ht]
\begin{center}
\includegraphics[width=6.25in]{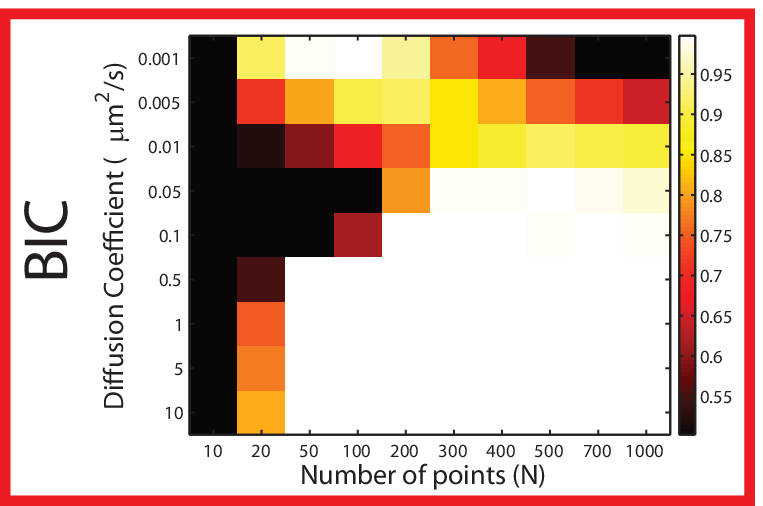}
\end{center}
\caption {
{\bf The BIC erroneously classifies the $4$th order potentials to be $2$nd order spring potentials.} The 2D plot shows the heat map of the percentage of trajectories classified to have a $2$nd order spring potential out of $300$ simulated $4$th order $(V=\alpha r^{4})$ trajectories per square for the BIC. The heat map has a threshold of $0.5$, which means that only cases where the information criterion falsely chooses 2nd order more than half of the time are non-black as indicated by the color scale. As mentioned earlier, the BIC cannot correctly attribute the $4$th order potential, but finds a $2$nd order spring-potential instead. Although this is clearly wrong, it can be exploited to build a two-step decision tree that can correctly distinguish all three cases using a mixture of BIC and AIC.}
\label{FigS5}
\end{figure}

\begin{figure}[!ht]
\begin{center}
\includegraphics[width=3.27in]{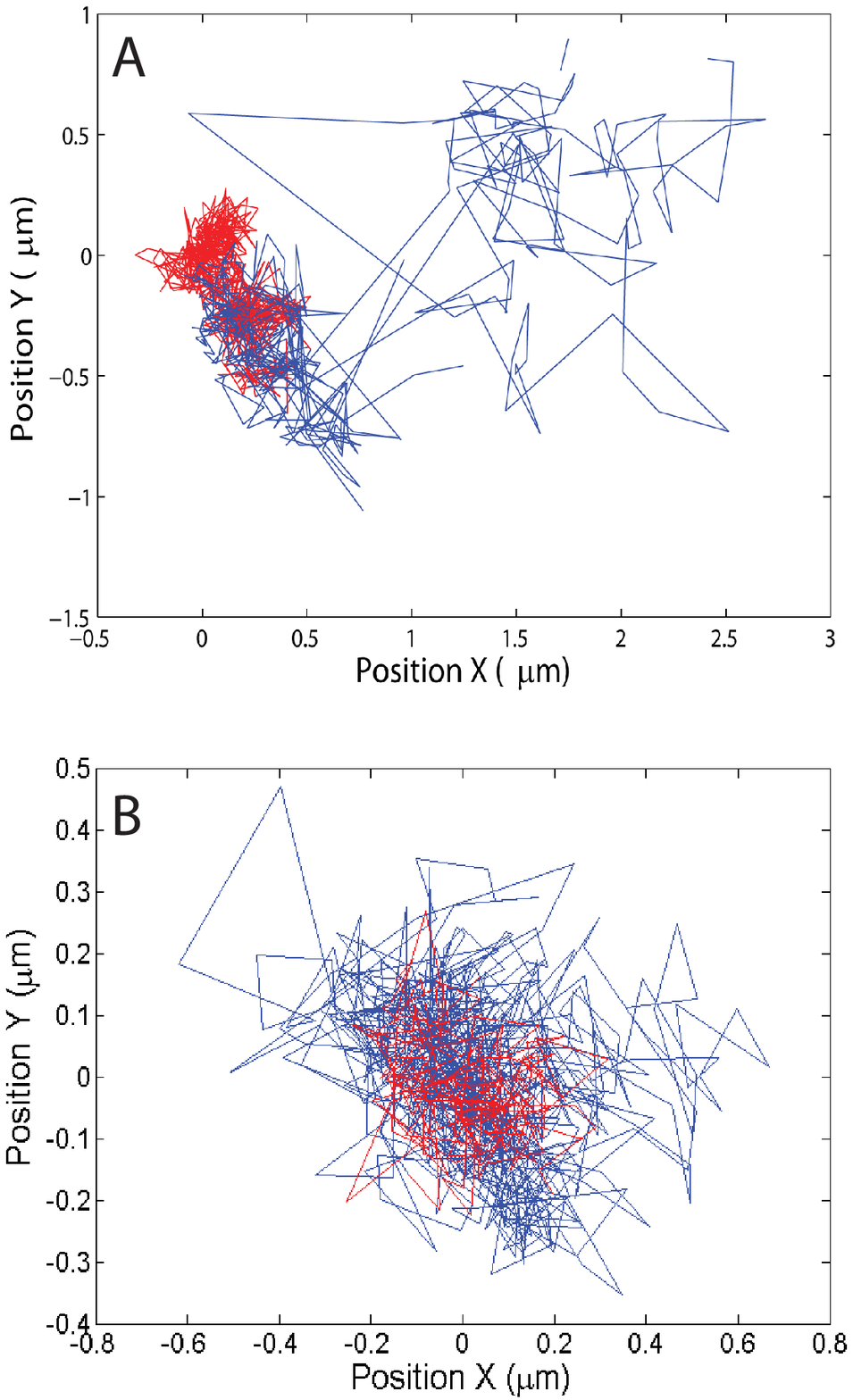}
\end{center}
\caption {
{\bf Inferring the mode of motion of single experimental CPɛT receptors during disaggregation of the confining domain.} Two different trajectories are shown in A and B. We use the decision tree to determine the mode of motion on a $51$ frame window that slides along the trajectory in time, while the cells are treated with cholesterol oxidase. The trajectory begins being confined by a spring-like potential (red). As the enzyme cholesterol oxidase oxidizes more cholesterol, the trajectory receptor becomes less confined and undergoes free Brownian motion (blue).}
\label{FigS6}
\end{figure}

\begin{figure}[!ht]
\begin{center}
\includegraphics[width=4.25in]{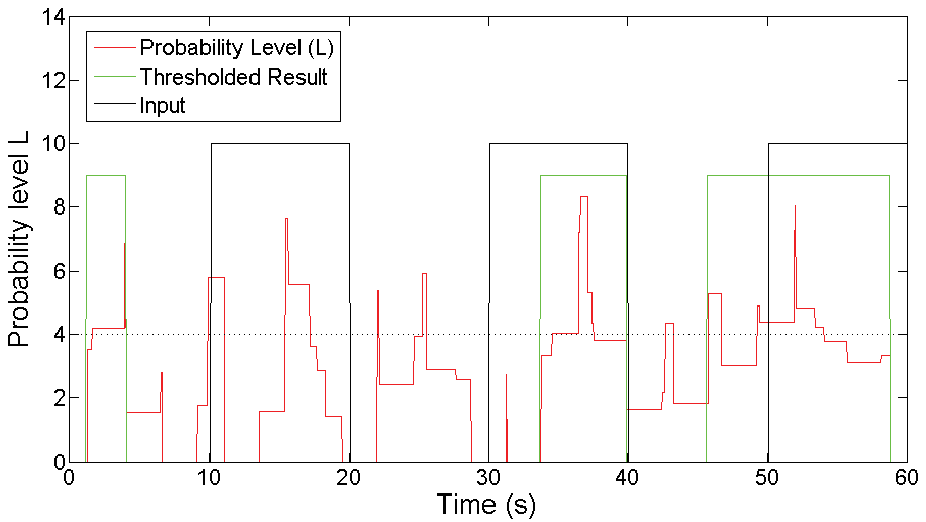}
\end{center}
\caption {
{\bf Residence time method for the detection of temporal lateral confinement.} A single trajectory has three different confinement zones in a second order potential. (Parameters: $D_{input}=0.1\,\mu m^{2}/s$, $t_{acq}=50\,ms$, $B_{r}=30\,nm$, $k=0.3\,pN/\mu m$). The three confined parts of the trajectory are shown in black where a non-zero value indicates confinement. Outside of the confined zones, the trajectory undergoes free Brownian motion. The probability level (\textit{L}) is shown in red along the trajectory. \textit{L} is filtered in magnitude with a cutoff $L_{c}=4$ (dotted line) and temporal $t_{c}=2.5\,s$. To qualify as a confinement zone, \textit{L} has to be larger than $L_{c}$ for a time greater than $t_{c}$. The parameters were optimized to detect the confinement zones without generating many false positives. The threshold result displaying the confinement zones is shown in green. The method could correctly determine two out of three confining zones and found one false confinement zone (false-positive).}
\label{FigS7}
\end{figure}

\clearpage

\section*{Source code}

Source codes available at: http://journals.plos.org/plosone/article?id=10.1371/journal.pone.0082799.\\

Source Code S8: Code in C (\textit{GenerateBrownianTraj.c}) to generate a trajectory undergoing free Brownian motion.\\

Source Code S9: Code in C (\textit{GenerateBrownianTrajin2ndOrderPot.c}) to generate a trajectory undergoing free Brownian motion confined in a 2nd order spring potential.\\

Source Code S10: Code in C (\textit{GenerateBrownianTrajin4thOrderPot.c}) to generate a trajectory undergoing free Brownian motion confined in a 4th order potential.\\

Source Code S11: Code in C (\textit{Model distinction in trajectory.c}) to calculate the information criteria for a given trajectory. This algorithm requires the files: \textit{nrutil.c} (Source code S12), \textit{nrutil.h} (Source code S13) and \textit{simplex.c} (Source code S14).\\

Source Code S12: Code in C (\textit{nrutil.c}) is required to run \textit{Model distinction in trajectory.c} (Source Code S11).\\

Source Code S13: Code in C (\textit{nrutil.h}) is required to run \textit{Model distinction in trajectory.c} (Source Code S11).\\

Source Code S14: Code in C (\textit{simplex.c}) is required to run \textit{Model distinction in trajectory.c} (Source Code S11).\\

\end{document}